\documentclass[twocolumn, nofootinbib, aps, prd, floats, floatfix, amsmath, amssymb, longbibliography, superscriptaddress, preprintnumbers]{revtex4-1} %
\usepackage[final]{graphicx}
\usepackage{hyperref}
\usepackage{amsmath}
\usepackage{bbm}
\usepackage{bm}
\usepackage{amsfonts}
\usepackage{amssymb}
\usepackage{latexsym}
\usepackage{graphicx}
\usepackage[english]{babel}
\usepackage{multirow}
\usepackage{float}
\usepackage{url}
\usepackage{slashed}
\usepackage{xcolor} 
\usepackage[utf8]{inputenc}
\usepackage{stmaryrd} 
\usepackage{enumitem}
\usepackage{hyperref}
\usepackage{cleveref}
\usepackage{siunitx}
\usepackage{verbatim}

\usepackage{amsthm}

\theoremstyle{remark}

\newcommand{\be}{\begin{equation}}
	\newcommand{\ee}{\end{equation}}
\newcommand{\ba}{\begin{array}}
	\newcommand{\ea}{\end{array}}
\newcommand{\bea}{\begin{eqnarray}}
	\newcommand{\eea}{\end{eqnarray}}

\renewcommand{\d}{\mathrm{d}}

\newcommand{\nn}{\nonumber}

\newcommand{\bE}{{\bf E}}
\newcommand{\bA}{{\bf A}}

\newcommand{\cC}{{\cal C}}
\newcommand{\besub}{\begin{subequations}}
	\newcommand{\eesub}{\end{subequations}}
\newcommand{\bs}[1]{\boldsymbol{#1}}

\newcommand{\eg}{e.g.,\ }	
\newcommand{\ie}{i.e.,\ }	

\newcommand{\kHz}{{\rm kHz}}

\newcommand{\GHz}{{\rm GHz}}
\newcommand{\beq}{\begin{equation} \begin{aligned}}
		\newcommand{\eeq}{\end{aligned} \end{equation}}

\definecolor{darkerblue}{rgb}{0.2,0.2,0.5}
\definecolor{seagreen}{rgb}{0.180392,0.545098,0.341176}
\definecolor{smagenta}{rgb}{0.5,0.145098,0.341176}
\definecolor{deepblue}{rgb}{0,0,1}

\begin{document}

	\title{Direct detection of dark photon dark matter using radio telescopes}
	
	\author{Haipeng An}
	\email{anhp@mail.tsinghua.edu.cn}
	\affiliation{Department of Physics, Tsinghua University, Beijing 100084, China}
	\affiliation{Center for High Energy Physics, Tsinghua University, Beijing 100084, China}
	\affiliation{Center for High Energy Physics, Peking University, Beijing 100871, China}	
	\affiliation{Frontier Science Center for Quantum Information, Beijing 100084, China}
	
	\author{Shuailiang Ge}
	\email{sge@pku.edu.cn}
    \affiliation{Center for High Energy Physics, Peking University, Beijing 100871, China}
	\affiliation{School of Physics and State Key Laboratory of Nuclear Physics and Technology, Peking University, Beijing 100871, China}

	\author{Wen-Qing Guo}
	\email{guowq@pmo.ac.cn}
	\affiliation{Key Laboratory of Dark Matter and Space Astronomy, Purple Mountain Observatory, Chinese Academy of Sciences, Nanjing 210033, China}
	\affiliation{School of Astronomy and Space Science, University of Science and Technology of China, Hefei, Anhui 230026, China}
	
	\author{Xiaoyuan Huang}
	\email{xyhuang@pmo.ac.cn}
	\affiliation{Key Laboratory of Dark Matter and Space Astronomy, Purple Mountain Observatory, Chinese Academy of Sciences, Nanjing 210033, China}
	\affiliation{School of Astronomy and Space Science, University of Science and Technology of China, Hefei, Anhui 230026, China}
	
	\author{Jia Liu}
	\email{jialiu@pku.edu.cn}
	\affiliation{School of Physics and State Key Laboratory of Nuclear Physics and Technology, Peking University, Beijing 100871, China}
	\affiliation{Center for High Energy Physics, Peking University, Beijing 100871, China}

	\author{Zhiyao Lu}
	\email{2000011457@stu.pku.edu.cn}
	\affiliation{School of Physics and State Key Laboratory of Nuclear Physics and Technology, Peking University, Beijing 100871, China}
	
	\begin{abstract}
		Dark photons can be the ultralight dark matter candidate, interacting with Standard Model particles via kinetic mixing. We propose to search for ultralight dark photon dark matter (DPDM) through the local absorption at different radio telescopes. The local DPDM can induce harmonic oscillations of electrons inside the antenna of radio telescopes. It leads to a monochromatic radio signal and can be recorded by telescope receivers. Using the observation data from the FAST telescope, the upper limit on the kinetic mixing can already reach $10^{-12}$ for DPDM oscillation frequencies at $1-1.5$ GHz, which is stronger than the cosmic microwave background constraint by about one order of magnitude. Furthermore, large-scale interferometric arrays like LOFAR and SKA1 telescopes can achieve extraordinary sensitivities for direct DPDM search from 10 MHz to 10 GHz. 
	\end{abstract}
	\maketitle
	
	\noindent \textit{\textbf{Introduction}}--
	Ultralight bosons are attractive dark matter (DM) candidates, including QCD axions, axion-like particles, dark photons, etc.~\cite{Essig:2013lka, Battaglieri:2017aum, ParticleDataGroup:2020ssz}. Dark photon mixed with photon through a marginal operator at low energy is one of the simplest extensions beyond the Standard Model of particle physics~\cite{Holdom:1985ag, Dienes:1996zr, Abel:2003ue, Abel:2006qt, Abel:2008ai, Goodsell:2009xc}. It can be a force mediator in the dark sector~\cite{Essig:2013lka, Fabbrichesi:2020wbt, Caputo:2021eaa} or a DM candidate itself~\cite{Redondo:2008ec, Nelson:2011sf, Arias:2012az, Graham:2015rva}.
	
	This work focuses on the dark photon dark matter (DPDM) with a mass, $m_{A'}$, comparable to the energy of radio frequency photons (20 kHz -- 300 GHz).
	Ultralight DPDM can be produced through inflationary fluctuations~\cite{Graham:2015rva, Ema:2019yrd, Kolb:2020fwh, Salehian:2020asa, Ahmed:2020fhc, Nakai:2020cfw, Nakayama:2020ikz, Kolb:2020fwh, Salehian:2020asa, Firouzjahi:2020whk, Bastero-Gil:2021wsf, Firouzjahi:2021lov, Sato:2022jya}, parametric resonances~\cite{Co:2018lka, Dror:2018pdh, Bastero-Gil:2018uel, Agrawal:2018vin, Co:2021rhi,  Nakayama:2021avl},  cosmic strings~\cite{Long:2019lwl}, and the non-minimal coupling enhanced misalignment~\cite{Nelson:2011sf, Arias:2012az, AlonsoAlvarez:2019cgw} with possible ghost instability \cite{Nakayama:2019rhg, Nakayama:2020rka}. Radio-frequency DPDM can be constrained indirectly by Cosmic Microwave Background (CMB) spectrum distortion~\cite{Arias:2012az, McDermott:2019lch, Arias:2012az, McDermott:2019lch, Caputo:2020bdy, Witte:2020rvb}
	and directly by haloscope experiments like TOKYO~\cite{Suzuki:2015vka, Suzuki:2015sza, Knirck:2018ojz, Tomita:2020usq}, FUNK \cite{FUNKExperiment:2020ofv}, DM pathfinder and Dark E-field \cite{Phipps:2019cqy, Godfrey:2021tvs}, SHUKET \cite{Brun:2019kak}, WISPDMX \cite{Nguyen:2019xuh}, SQuAD \cite{Dixit:2020ymh},
	and recent experiments~\cite{
		Cervantes:2022gtv, Ramanathan:2022egk, Cervantes:2022yzp, DOSUE-RR:2022ise}.
	Axion haloscope search results~\cite{Asztalos:2009yp, ADMX:2018gho, ADMX:2019uok, ADMX:2018ogs, HAYSTAC:2018rwy, HAYSTAC:2020kwv, Alesini:2020vny, Lee:2020cfj, Jeong:2020cwz, CAPP:2020utb,
		ADMX:2021nhd, Crisosto:2019fcj, Lee:2022mnc, Kim:2022hmg, Yi:2022fmn, Adair:2022rtw, HAYSTAC:2023cam, Quiskamp:2022pks, Alesini:2019ajt, Alesini:2022lnp, TASEH:2022vvu} 
	can be interpreted to DPDM limits~\cite{Caputo:2021eaa,DPLimits}, but some searches relying on the magnetic veto, e.g. RBF~\cite{ DePanfilis:1987dk} and UF~\cite{Hagmann:1990tj}, cannot be translated into DPDM limits~\cite{Gelmini:2020kcu, Caputo:2021eaa}.
	Proposals and future experiments to search for DPDM include plasma haloscopes~\cite{Lawson:2019brd, Gelmini:2020kcu}, Dark E-field \cite{Godfrey:2021tvs}, DM-Radio~\cite{Parker:2013fxa, Chaudhuri:2014dla, 7750582, Chaudhuri:2018rqn}, MADMAX~\cite{Caldwell:2016dcw}, and solar radio observations~\cite{An:2020jmf, An:2023wij}.

	One category of broadband haloscope experiments uses a dish reflector to look for dark photons~\cite{Horns:2012jf, Jaeckel:2013eha, Jaeckel:2015kea}. The original proposal uses a spherical reflector to convert $A' \to \gamma$, and the monochromatic photons with energy $m_{A'} c^2$ are emitted perpendicular to the surface, thus focusing on the spherical center. This method has been applied to room-sized experiments \cite{Suzuki:2015vka, Suzuki:2015sza, Knirck:2018ojz, Tomita:2020usq,  Godfrey:2021tvs, Brun:2019kak, FUNKExperiment:2020ofv}, with variations using plane/parabolic reflectors or dipole antenna placed in a shielded room. 
	
	In this work, we propose to use existing and future radio telescopes to search for DPDM directly. With huge effective areas and great detectors, the sensitivities of large-scale radio telescopes can surpass current astrophysics bounds on radio-frequency DPDM by several orders of magnitude. We perform two types of studies: one exploits a single large dish antenna to convert dark photons into radio signals; the other uses antenna arrays forming interferometer pairs to receive radio signals, taking advantage of the long DPDM coherence. 
	
\begin{figure}
		\centering
		\includegraphics[width=1. \columnwidth]{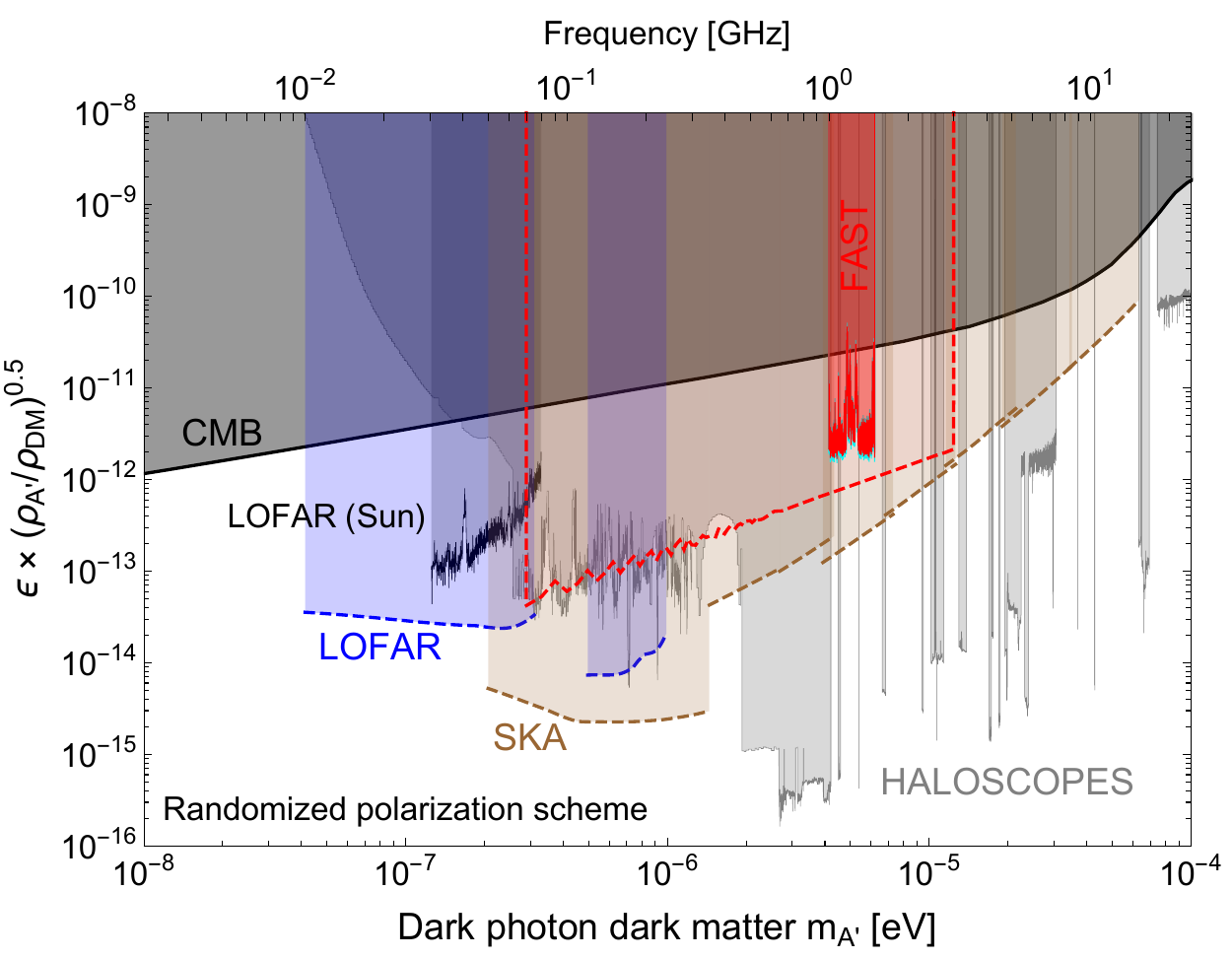} 
		\caption{Constraints and projected sensitivities on the kinetic mixing $\epsilon$ between DPDM and photon in the randomized polarization scheme. The $95\%$ confidence level (C.L.) exclusion limit for DPDM using the FAST data is represented by the solid red curve with an $\mathcal{O}(10\%)$ uncertainty in cyan, while the dashed red curve indicates its future sensitivity projection.
			The blue and brown dashed curves show the future sensitivity projections of LOFAR and SKA1 interferometric array telescopes. 
			The existing limits are from CMB constraints \cite{Arias:2012az, McDermott:2019lch, Arias:2012az, McDermott:2019lch, Caputo:2020bdy, Witte:2020rvb}, 
			solar radio observations \cite{An:2020jmf, An:2023wij}, 
			various haloscope searches \cite{
				Godfrey:2021tvs, Brun:2019kak, Nguyen:2019xuh, Dixit:2020ymh, 
				Cervantes:2022gtv, Ramanathan:2022egk, Cervantes:2022yzp, DOSUE-RR:2022ise}, 
			and axion experiments 
			\cite{Asztalos:2009yp, ADMX:2018gho, ADMX:2019uok, ADMX:2018ogs, HAYSTAC:2018rwy, HAYSTAC:2020kwv, Alesini:2020vny, Lee:2020cfj, Jeong:2020cwz, CAPP:2020utb,
				ADMX:2021nhd, Crisosto:2019fcj, Lee:2022mnc, Kim:2022hmg, Yi:2022fmn, Adair:2022rtw, HAYSTAC:2023cam, Quiskamp:2022pks, Alesini:2019ajt, Alesini:2022lnp, TASEH:2022vvu} translated to  randomized polarization scheme  \cite{Caputo:2021eaa, DPLimits}. 
		}	\label{fig:fin}
	\end{figure}

	Fig.~\ref{fig:fin} summarizes our main results. The FAST data excludes the region surrounded by the solid red curve. The dashed red, blue, and brown curves show the projected sensitivities of FAST, LOFAR, and SKA1~\cite{footnote} telescopes, assuming one-hour observation. For comparison, CMB and haloscopes constraints are shown by the black and gray shaded regions, respectively. The results show that large radio telescopes can play an essential and complementary role in DPDM searches. 
	\\

	\noindent \textit{\textbf{Model}}--	
	We consider the dark photon Lagrangian 
	\begin{equation}
		\mathcal{L} =  - \frac{1}{4} F'_{\mu\nu} F'^{\mu\nu}  + \frac{1}{2} m_{A'}^2 A'_{\mu} A'^{\mu} -\frac{1}{2} \epsilon  F_{\mu \nu} F'^{\mu \nu} \ .
	\end{equation}
	$F'$ and $F$ are dark photons and SM photons field strength; $\epsilon$ is the kinetic mixing. 
	After appropriate rotation and redefinition, one can eliminate the kinetic mixing term and arrive at the interaction Lagrangian for $A'$, the SM photon $A$, and the electromagnetic current $j^\mu_{\rm em}$,
	\begin{equation}\label{eq:interaction}
		\mathcal{L}_{\rm int} =  e j^\mu_{\rm em} \left(A_\mu  - \epsilon   A'_\mu \right).
	\end{equation}
	$e$ is the electromagnetic coupling. 
	Therefore, free electrons in telescope antennas will be accelerated by the DPDM electric field, $\bE' = - \dot{\bA}' - \nabla A'^0$, and then produce EM equivalent signals. 
	
	Since the local DM velocity is about $10^{-3}c$, where $c$ is the speed of light, $\bE'$ oscillates with a nearly monochromatic frequency, $f\approx m_{A'}/2\pi$. Therefore, radio telescopes will detect a monochromatic radio signal, broadening the center value of about $10^{-6}$. The DPDM wavelength is about $10^3 c/f$, $10^3$ times the same-frequency EM wavelength. Next, we analyze DPDM signals for the dipole antenna, dish antenna, and antenna arrays.   
	\\

	\noindent \textit{\textbf{Response of the dipole antenna}}--
	A dipole antenna usually comprises conductive elements like metal wires or rods. Considering a linear dipole antenna of length $ \ell$ lying on the horizontal plane observing a radio signal from the zenith direction with frequency $f$, it will detect an oscillating electric field 
	\begin{equation}
		\label{eq:EEM}
		E_{\rm EM} = E_0 \cos\psi \cos \left(2\pi f t  - {\bf k}\cdot {\bf x}\right)  \ .
	\end{equation}
	$E_0$ is the amplitude, ${\bf k}$ is the wave number, and $\psi$ is the angle between the electric field and the antenna rod. $\ell$ is usually around half of the EM wavelength designed to detect.
	However, the DPDM wave number ${\bf k}'$ is about ${\cal O}(10^{-3})$ times smaller than ${\bf k}$ due to the small DM velocity. Therefore, according to Eq.~(\ref{eq:interaction}), the antenna will register an equivalent electric field,
	\begin{align}
		\label{eq:EDP}
		E_{\rm EM}^{\rm eqv} &= \epsilon E'_0 \cos\psi' \cos{ \left(2\pi f t - {\bf k}' \cdot {\bf x} \right)} , \nonumber \\
		&\simeq
		\epsilon E'_0 \cos\psi' \cos{ \left(2\pi f t \right)} \ .
	\end{align}
	$E'_0$ is the amplitude of the dark electric field. $\psi'$ is the angle between the dark electric field and the antenna rod.

	Thus, typical dipole antennas respond to EM and DPDM fields differently, mainly by factors of $\epsilon$ and the polarization angle. Additionally, for the DPDM case, the antenna can always be seen as a \textit{short} dipole antenna since ${k'}\ell \ll 1$ for proposed frequencies, modifying the antenna efficiency by an $\mathcal{O}(1)$ number. Therefore, for general dipole antennas, one can define a DPDM-induced equivalent EM flux density,
	\begin{equation}
		\label{eq:dipole}
		I_{\rm dipole}^{\rm eqv} \equiv {\cC}_{\rm dipole} \epsilon^2  \langle \bE'^2 \rangle = {\cC}_{\rm dipole}\epsilon^2 \rho_{\rm DM} \ .
	\end{equation}
	$\rho_{\rm DM} = 0.3~{\rm GeV/cm^3}$ is the conservative local DM energy density \cite{deSalas:2019pee,deSalas:2020hbh}. ${\cal C}_{\rm dipole}$ is an ${\cal O}(1)$ numerical factor. For telescopes like LOFAR and SKA1-Low, detailed antenna designs are needed to simulate the exact values of ${\cal C}_{\rm dipole}$, which is beyond the scope of the present work. Instead, we prove that ${\cal C}_{\rm dipole} \ge 1$ for the antenna with linear dipole configuration, showing that the DPDM signal gains enhancement over the EM signal in Sec.~\ref{sec:C-for-dipole} of the supplemental material (SUPP) \cite{chatterjee1996antenna}. In this work, we \textit{conservatively} assume ${\cal C}_{\rm dipole} = 1$ to estimate the potential sensitivity of LOFAR and SKA1-Low. 
	\\

	\noindent \textit{\textbf{Response of the dish antenna}}--
	Some large radio telescopes are constructed as dish antennas like FAST~\cite{2011IJMPD..20..989N} or dish antenna arrays like MeerKAT \cite{2016mks..confE.....} and SKA1-Mid \cite{SKA1-MID-configuration}. A dish antenna usually comprises a parabolic reflector with the feed receiving reflected EM wave at the focus. Dishes are commonly made of metal plates. According to Eq.~(\ref{eq:interaction}), DPDM causes free electrons on metal plates to oscillate. Thus, each area unit can be seen as an oscillating dipole emitting EM waves with the same frequency as DPDM. Then, the feed signal is the integration over the dipole units. In Sec.~\ref{sec:dishantenna-1} of SUPP, we show that the induced dipole with area $dS$ is
	\begin{equation}
		d{\bf p} = 2 \epsilon {\bf A}'_\parallel d S \ .
	\end{equation}
	${\bf A}'_\parallel$ is the projection of ${\bA}'$ on $dS$. 
	Then, the EM field at position ${\bf r}$ can be obtained by summing up area units,
	\begin{align}
		{\bf B} = - \frac{\epsilon m_{A'}^2}{2 \pi} \int dS_1 {\bf A'_{\parallel}} \times ({\bf r} - {\bf r}_1) \frac{e^{i  m_{A'} |{\bf r} - {\bf r}_1|}}{|{\bf r} - {\bf r}_1|^2}.
	\end{align}
	The electric field ${\bf E}$ can be calculated using ${\bf B} $. EM phase at each dipole unit is determined by the DPDM wavelength, $\lambda'$, different from the phase induced by parallel EM waves from distant stars. Therefore, the EM wave generated by DPDM will not focus on the antenna feed. For a single filled-aperture telescope like FAST, its diameter can be comparable to $\lambda'$. Thus, numerical simulation is necessary to calculate the induced EM flux into the feed. However, for dish antenna arrays like MeerKAT and SKA1-Mid, each dish's diameter is much smaller than $\lambda'$. Therefore, each dish's dipole units $d{\bf p}$ oscillate in phase. 
	
	Due to the continuous boundary condition for the electric field parallel to the metal surface, we have $\bE_\parallel = \epsilon \bE_\parallel'$ right outside the metal surface and the perpendicular component $|\bE_\perp|/|\bE_\parallel| \sim (f \lambda')^{-1} \approx 10^{-3}$. In Sec.~\ref{sec:dishantenna-1} of SUPP, detailed calculation shows that the {\it reflected} EM wave propagates nearly perpendicular to the surface of the metal plate. Right on top of the reflector surface, its energy density can be estimated as $\epsilon^2 |\bE'|^2 \cos^2\theta$, where $\theta$ is the angle between $\bE'$ and the reflector plate.     
	
	Since the DPDM-induced EM wave is not focusing, its flux into the feed is much smaller than the total reflected flux. The parabolic antenna feed size is usually around the EM wavelength $\lambda$ to optimize the absorption, so the reduction factor is roughly, $\lambda^2/{\cal A}$, the ratio between feed and reflector areas.
	Therefore, compared to the EM signal from distant sources, the DPDM-induced equivalent EM flux density can be written as 
	\begin{equation}
		\label{eq:dish}
		I^{\rm eqv}_{\rm dish} = {\cal C}_{\rm dish} \epsilon^2 \langle \bE'^2 \rangle \times \frac{\lambda^2}{{\cal A}} 
		= {\cal C}_{\rm dish} \epsilon^2 \rho_{\rm DM} \frac{\lambda^2}{{\cal A}} \ .
	\end{equation}
	${\cal C}_{\rm dish}$ is an ${\cal O}(1)$ numerical factor determined by the detailed antenna design. Numerical calculations of ${\cal C}_{\rm dish}$ are performed by averaging all possible $A'$ polarization, denoted as the randomized polarization scheme. 
	Results for FAST and SKA1-Mid are shown in Sec.~\ref{sec:C-dish-numeric} of SUPP. 
	\\
	
	\noindent \textit{\textbf{Sensitivities of antenna arrays}}--	
	Radio telescopes using radio interferometry techniques can effectively enlarge the effective area and get better sensitivities on faint signals. The basic observation unit for radio interferometer array is the antenna pair~\cite{thompson2001}. Let $V_m(t)$ and $V_n(t)$ be the signal measured by the $m$-th and $n$-th antenna, then up to amplification factors, the pair's output signal is 
	\begin{equation}
		r_{mn} =\langle V_m(t) V^{*}_n(t) \rangle \ .
	\end{equation}
	$\langle \cdots \rangle$ means the time average. $V_m$ and $V_n$ can be seen as the voltage measured by antennas, proportional to the electric field. Therefore, the correlator $r_{mn}$ is proportional to the EM flux density~\cite{thompson2001}.  
	A telescope composed of $N$ antennas has $N(N-1)/2$ independent pairs. The combined signal increases as $N(N-1)/2$, whereas the noise goes like $[N(N-1)/2]^{1/2}$. Thus, the signal-over-noise-ratio increases as $[N(N-1)/2]^{1/2} \approx N/\sqrt{2}$. 
	
	For normal EM signals, the minimum detectable spectral flux density of a radio telescope is 
	\begin{equation}
		\label{eq:Smin}
		S_{\rm min} = \frac{\rm SEFD}{\eta_s \sqrt{n_{\rm pol} {\cal B} t_{\rm obs}}} \ .
	\end{equation}
	$n_{\rm pol} =2$ is the number of polarizations, $\eta_s$ is the system efficiency, $t_{\rm obs}$ is the observation time, ${\cal B}$ is the bandwidth, and SEFD is the system equivalent (spectral) flux density,
	\begin{equation}
		{\rm SEFD} = \frac{2k_B  T_{\rm sys}}{A_{\rm eff}} \ .
	\end{equation}
	$T_{\rm sys}$ is the antenna system temperature. $A_{\rm eff}$ is the antenna array's effective area, increasing with the number of antennas, $N$. 
	
	For the DPDM-induced signal, the correlation length is determined by its wavelength, $\lambda'$, beyond which the DPDM oscillation is out of phase; thus, the correlation is suppressed. For two antennas with distance $d_{mn}$, the correlation signal is suppressed by
	\begin{equation}\label{eq:Smn}
		{\cal S}_{mn} \approx \exp(- m_{A'}^2 v_0^2 d_{mn}^2/8) \ .
	\end{equation}
	$v_0 \approx 235 ~{\rm km /s}$ is the most probable velocity in the Standard Halo Model \cite{McMillan:2009yr, Bovy:2009dr}. The detailed derivation uses truncated Maxwellian distribution, as shown in Sec.~\ref{sec:correlation-of-antenna} of SUPP~\cite{Foster:2017hbq, OHare:2018trr, Evans:2018bqy, OHare:2019qxc, Foster:2020fln}, consistent with Ref.~\cite{Derevianko:2016vpm}. 
	
	Therefore, for an antenna array composed of $N$ antennas, the DPDM-induced equivalent EM flux density is
	\begin{equation}
		I^{\rm eqv}_{\rm array} = {\cal S}_{\rm eff} I^{\rm eqv}_{\rm single} \ ,
	\end{equation}
	where 
	\begin{equation}
		{\cal S}_{\rm eff} = \frac{2}{N(N-1)} \sum_{m=2}^{N}\sum_{n=1}^{m} {\cal S}_{mn} \ ,
	\end{equation}
	is the suppression factor. $I^{\rm eqv}_{\rm single}$ is the DPDM-induced EM flux density for an individual antenna, given by (\ref{eq:dipole}) for dipole antenna and ({\ref{eq:dish}}) for dish antenna. For dipole array telescopes like LOFAR and SKA1-Low, the antennas first form stations, which are further organized into a large interferometer. Since each station's size is much smaller than $\lambda'$, we neglect the suppression within a station. Therefore, the suppression factor becomes
	\begin{equation}
		{\cal S}_{\rm eff} = \frac{2}{N_{\rm stat}(N_{\rm stat}-1)} \sum_{m=2}^{N_{\rm stat}}\sum_{n=1}^{m} {\cal S}_{mn} \ .
	\end{equation}
	$N_{\rm stat}$ is the number of stations. $d_{mn}$ is the distance between the $m$-th and $n$-th stations. 
	
	Next, we will use the criterion
	\begin{equation}
		I^{\rm eqv}_{\rm array}/ \mathcal{B} > S_{\rm min} 
		\label{eq:signalcri}
	\end{equation}
	to estimate the projected sensitivities of LOFAR and SKA1 arrays for DPDM. 
	\\

	\noindent \textit{\textbf{Constraints from FAST observation data}}--
	FAST is currently the largest filled-aperture radio telescope. Its designed total bandwidth is from 70 MHz to 3 GHz with the current frequency resolution ${\cal B} = 7.63$ kHz and designed sensitivity SEFD$^{-1}$ = 2000 m$^2/$K~\cite{2011IJMPD..20..989N, 2020Innov...100053Q}. During observation, a 300-meter aperture instantaneous paraboloid is formed to reflect and focus the EM wave into the feed. The DPDM-induced EM wave is not focusing and therefore suffers from the suppression factor, $\lambda^2/{\cal A}$; see~(\ref{eq:dish}). The simulation of the factor ${\cal C}$ for FAST at different frequencies is detailed in Sec.~\ref{sec:C-FAST-simulation} of SUPP, from which we can calculate the DPDM-induced EM spectral flux density detected by FAST,
	\begin{equation}
		\label{eq:S_limit_eqv}
		S^{\rm eqv}_{\rm FAST}(f) \equiv 
		\frac{I^{\rm eqv}_{\rm FAST}}{\mathcal{B}}
		\approx 4.6\times 10^{-6}  \epsilon^2 
		\frac{{\cal C}_{\rm FAST}(f)}{{\cal C}_{\rm FAST}(1{\rm GHz})}
		\frac{{\rm W}}{{\rm m^2~Hz}}.
	\end{equation}
	Requiring $S^{\rm eqv}_{\rm FAST} > S_{\rm min}$, we can calculate the sensitivity for the FAST telescope.

	Apart from the simulation, we use the 19-beam L-band (1$-$1.5~GHz) observation data from FAST to set upper limits for DPDM. The observation was conducted on December 14, 2020, lasting 110 minutes.
	A time series of the signal is recorded for each frequency bin. We use the noise diode temperature to calibrate data and convert the signal to the EM spectral flux density by pre-measured antenna gain. DPDM induces a time-independent line spectrum signal, whereas most noise sources have transient features and can be reduced by data filtering processes~\cite{Foster:2022fxn}. Our data filtering process is detailed in Sec.~\ref{sec:FAST-data} of SUPP~\cite{Jiang:2019rnj, 2020RAA....20...64J, Foster:2022fxn, Cowan:2010js, Arias:2012az}. 
	
\begin{figure}
		\centering
		\includegraphics[width=1\linewidth]{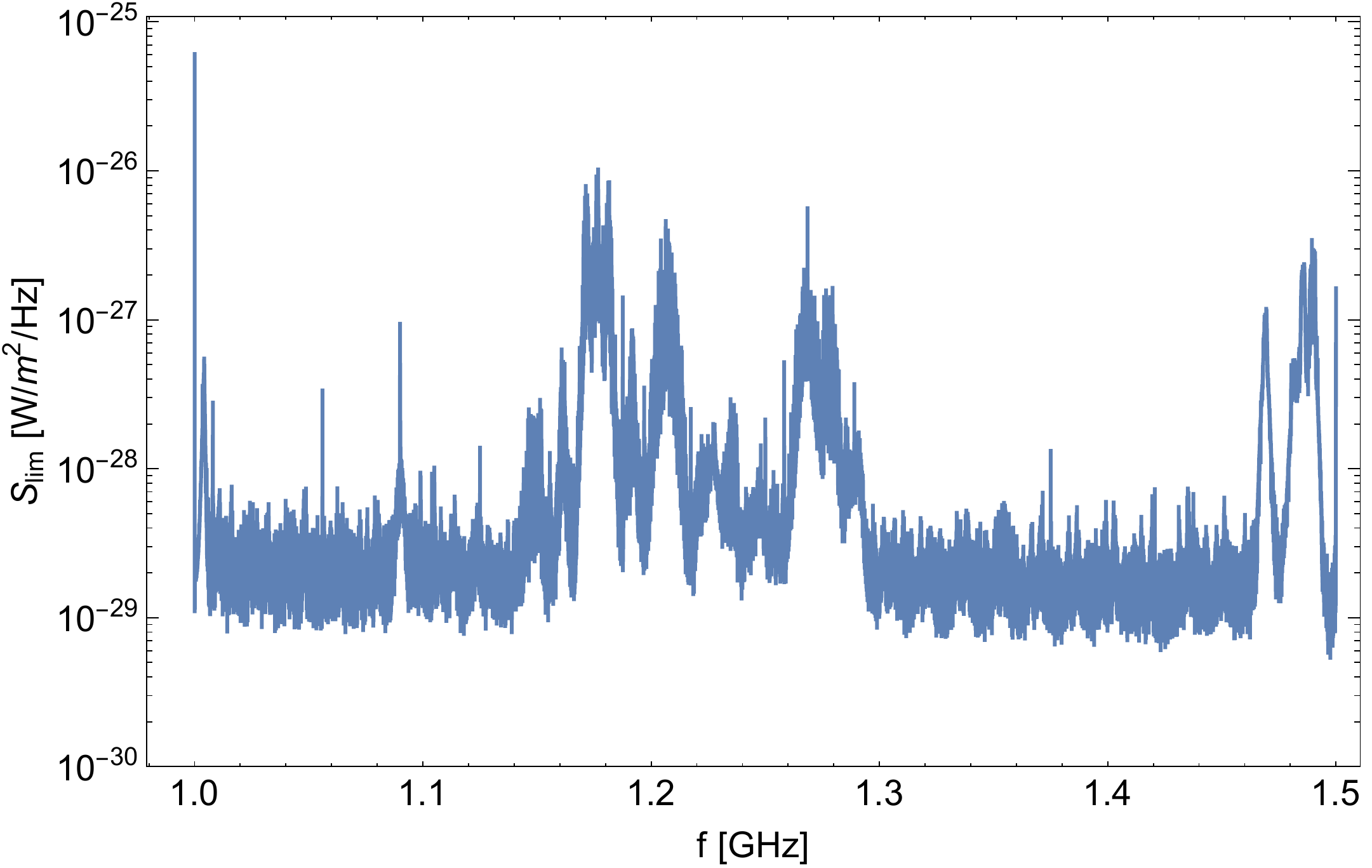}
		\caption{Model-independent 95\% C.L. upper limits on a constant monochromatic signal from FAST data in 1$-$1.5~GHz. It shows the strongest limit from the 19 beams at each frequency bin.}
		\label{fig:S_limit}
	\end{figure}

	After data filtering, for each frequency bin, $i$, we obtain the average measured spectral flux density, $\bar{O}_i$, and the statistic uncertainty, $\sigma_{\bar{O}_i}$. We then use a polynomial function to locally model the background around the selected frequency bin $i_0$ with the help of its neighboring frequency bins. The systematic uncertainty is estimated by the data deviation to the background fit. Next, we assume a dark photon signal with the strength $S$ existing at bin $i_0$, and a likelihood function $L$ can be built between data and background function with $S$ incorporated. Coefficients of the background polynomial function are treated as nuisance parameters. Following the likelihood-based statistical method~\cite{Cowan:2010js}, we compute the ratio $\lambda_S$ between the conditional maximized-likelihood (e.g., only varying the nuisance parameters to maximize $L$ while keeping $S$ fixed) and the unconditional maximized-likelihood (e.g., varying both the nuisance parameters and $S$ to maximize $L$). Then the test statistic, $-2\ln\lambda_S$, follows the half-$\chi^2$ distribution~\cite{Cowan:2010js}. Thus, we obtain the $95\%$ C.L. upper limit, $S_{\rm lim}$, for a constant monochromatic signal, shown in Fig.~\ref{fig:S_limit}.
	
	Upper limits on the mixing parameter $\epsilon$ are obtained via $S_{\rm lim} = S^{\rm eqv}_{\rm FAST}$. All 19 beams give similar constraints as expected. We choose the strongest limit among the 19 constraints for each frequency bin as the final result, shown in Fig.~\ref{fig:fin}. The upper limits can reach $\epsilon\sim 10^{-12}$ in 1$-$1.5~GHz, about one order-of-magnitude better than the existing constraint from CMB measurement~\cite{Arias:2012az}.
	We emphasize that every single frequency between 1$-$1.5~GHz is constrained by the real data without any extrapolation. 
	We also explore the rare case where the DPDM signal falls into two bins due to its broadening. The sensitivity calculation is similar but with a doubled data bandwidth.
	More details about the FAST original data, filtering processing, statistical methods, and numerical calculations are given in Sec.~\ref{sec:FAST-data} of SUPP. 
	\\

	\noindent \textit{\textbf{Sensitivities of LOFAR and SKA1}}--
	LOFAR is currently the largest radio telescope operating at the lowest frequencies ($10-240$~MHz), containing low-band antennas (LBAs) and high-band antennas (HBAs). LOFAR antennas are grouped into 24 remote stations, each with a core size smaller than 2 kilometers. DPDM wavelength within the LOFAR frequency range is $1.2-30$~km. Therefore, we propose to use the core stations to search for DPDM. Station positions and relevant parameters can be found in Ref.~\cite{vanHaarlem:2013dsa}. The minimal frequency resolution, ${\cal B}_{\rm min}$, of LOFAR is about 700 Hz~\cite{vanHaarlem:2013dsa}. 
	
	SKA1 continuously covers 50 MHz$-$20 GHz, including SKA1-Low and SKA1-Mid telescopes. SKA1-Low has 131,072 dipole-like antennas grouped into 512 stations, covering 50$-$350 MHz, with ${\cal B}_{\rm min} = 1$~kHz. Station positions and relevant parameters can be found in Ref.~\cite{SKA1-LOW-configuration,SKA1-Baseline-2}.  SKA1-Mid contains 133 SKA1 15-meter diameter and 64 MeerKAT 13.5-meter diameter dish-antennas. Therefore, its sensitivity on DPDM suffers from the additional suppression factor, $\lambda^2/{\cal A}$; see Eq.~(\ref{eq:dish}). SKA1-Mid has five bands, and the sensitivity and frequency range can be found in \cite{SKA1-Baseline-3} and dish locations in \cite{SKA1-MID-configuration}. SKA1-Mid achieves ${\cal B}_{\rm min} = 200$ Hz smaller than the DPDM natural width. Therefore, to calculate its DPDM sensitivity, we use the natural width, ${\cal B} = 10^{-6} f$.  
	
\begin{figure}
		\centering
		\includegraphics[height=2.5in]{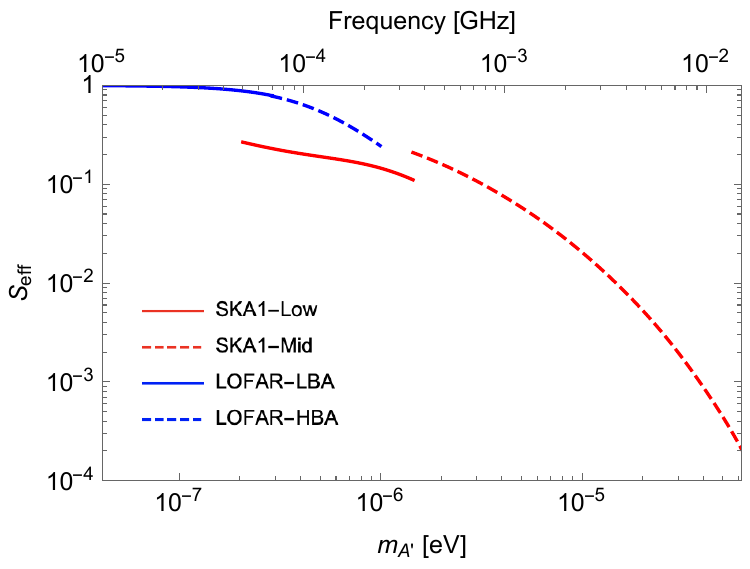}
		\caption{Suppression factor ${\cal S}_{\rm eff}$ for SKA1 and LOFAR (only core stations) arrays in the interferometry.}\label{fig:Ceff}
	\end{figure}

	The suppression factor ${\cal S}_{\rm eff}$ for DPDM signal using LOFAR and SKA1 arrays as interferometry are shown as the blue and red curves in Fig.~\ref{fig:Ceff}, respectively. LOFAR is less suppressed than SKA1 due to lower frequency, thus longer DPDM coherent wavelength and smaller separation between stations.
	
	Following Eq.~(\ref{eq:signalcri}), projected sensitivities on $\epsilon$ for LOFAR and SKA1 are shown in Fig.~\ref{fig:fin}. LOFAR can cover a frequency down to 10 MHz, complementary to Haloscope searches.
	SKA1 shows competitive sensitivity for higher frequencies as a broadband search compared to resonant cavity searches. 
	\\

	\noindent \textit{\textbf{Summary and Outlook}}--		
	The radio telescopes' antennas can convert the DPDM field into an ordinary EM wave. We have analyzed the sensitivities of the commonly used dipole and parabolic dish antennas. We found that the parabolic one has a significant suppression factor for the DPDM-induced equivalent EM flux. For antenna arrays like LOFAR and SKA1, due to the sizeable coherent length of DPDM, the interferometry technique in radio astronomy can enhance the sensitivity. 
	
	We have used FAST observational data to set limits for DPDM. The result is encouraging that for 1$-$1.5~GHz, the limit $\epsilon\sim 10^{-12}$ is one order of magnitude stronger than the CMB constraint. 
	We have projected the sensitivities for FAST, LOFAR, and SKA1 telescopes and found that compared to room-sized haloscope experiments, they are competing and complementary in searching for DPDM directly.

	The DPDM can directly interact with electrons through (\ref{eq:interaction}), inducing a signal in the feed. As detailed in SUPP, the signal induced from the reflector studied in this work is about four times larger than the direct feed signal, due to geometric reasons. 
	However, the feed shape is complex, making it difficult to calculate the direct contribution accurately. 
	The interference between the reflector and feed, along with the direct signal, may result in an $\mathcal{O}(10\%)$ uncertainty for the FAST limits in Fig.~\ref{fig:fin}.
	Furthermore, the FAST sensitivity could be significantly improved if one can raise the feed to higher locations as shown in SUPP.
	
	Dark photon mass can be generated through the Higgs mechanism or the St$\ddot{\rm u}$ckelberg mechanism. For the Higgsed case, the sub-keV dark photon is constrained to $\epsilon e_D < 10^{-14}$ by the stellar lifetime. $e_D$ is the dark U(1) gauge coupling. This assumes the dark Higgs has a dark charge of one and a mass below keV~\cite{An:2013yua}. Fig.~\ref{fig:fin} demonstrates that the proposed radio search complements the stellar constraint for small $e_D$ cases. For the St$\ddot{\rm u}$ckelberg case, the UV cutoff of dark photon model is constrained by the weak gravity conjecture~\cite{Reece:2018zvv, Montero:2022jrc}. Although some production mechanisms for radio DPDM, like inflation-induced DPDM~\cite{Graham:2015rva}, are no longer favored by certain constraints~\cite{Reece:2018zvv}, evading these constraints is possible by further developing the models~\cite{Craig:2018yld}. Therefore, a radio DPDM search could provide insights into DPDM production mechanisms.

	\begin{acknowledgments}
		\noindent \textit{\textbf{Acknowledgment}}--
		The authors would like to thank Peng Jiang, Yidong Xu, Jinglong Yu, Qiang Yuan and Yanxi Zhang for helpful discussions. This work made use of the data from FAST (Five-hundred-meter Aperture Spherical radio Telescope).  FAST is a Chinese national mega-science facility, operated by National Astronomical Observatories, Chinese Academy of Sciences.
		The work of HA is supported in part by the National Key R$\&$D Program of China under Grant No. 2021YFC2203100 and 2017YFA0402204, the NSFC under Grant No. 11975134, and the Tsinghua University Initiative Scientific Research Program.
		The work of SG is supported by NSFC under Grant No. 12247147, the International Postdoctoral Exchange Fellowship Program, and the Boya Postdoctoral Fellowship of Peking University. The work of XH is supported by the Chinese Academy of Sciences, and the Program for Innovative Talents and Entrepreneur in Jiangsu.
		The work of JL is supported by NSFC under Grant No. 12075005, 12235001 and by Peking University under startup Grant No. 7101502458. 
	\end{acknowledgments}
	
	
	\clearpage
	
	\setcounter{equation}{0}
	\setcounter{figure}{0}
	\setcounter{table}{0}
	\setcounter{section}{0}
	\setcounter{page}{1}
	\makeatletter
	\renewcommand{\theequation}{S\arabic{equation}}
	\renewcommand{\thefigure}{S\arabic{figure}}
	\renewcommand{\thetable}{S\arabic{table}}

\onecolumngrid
\begin{center}
	\textbf{\large Direct detection of dark photon dark matter using radio telescopes}\\[0.3cm]
	\vspace{0.05in}
\end{center}

 In the Appendix, we show the detailed calculation for the response of the dish antennas. Additionally, we discuss the response of dipole antennas and provide proof that the response factor is larger than one. Thirdly, we derive the correlation factor of two distant antennas for the sensitivity calculation of dipole antenna array telescopes. Lastly, we provide a detailed analysis of FAST observational data, including data processing, background fit, and likelihood-based statistical tests. 
 \\[0.3cm]
	
\twocolumngrid	
	\section{Response calculation for the dish antenna} \label{sec:dishantenna-1}
	The full Lagrangian for the kinetic mixing dark photon is
	\begin{align}
		\mathcal{L}= &-\frac{1}{4}F_{\mu\nu}F^{\mu\nu}-\frac{1}{4}F'_{\mu\nu}F'^{\mu\nu}+\frac{1}{2}m_{A'}^2A'_{\mu}A'^{\mu}\nn\\
		&-\epsilon eA'_\mu j^{\mu}_{\rm em}+eA_\mu j^{\mu}_{\rm em}\label{equ:lagrangian}.
	\end{align}
	By using variational principal, one obtains the Maxwell equations for the dark photon field,
	\begin{align}
		&\nabla \cdot \bs{E'}=-\epsilon \rho -m_{A'}^2 A'^0 \label{equ:divDarkE} ,\\
		&\nabla\cdot\bs{B'}=0\label{equ:divDarkB} ,\\
		&\nabla \times \bs{E'}+\frac{\partial \bs{B'}}{\partial t}=0\label{equ:rotDarkE} ,\\
		&\nabla \times \bs{B'}-\frac{\partial \bs{E'}}{\partial t}=-\epsilon \bs{J}-m_{A'}^2\bs{A'}\label{equ:rotDarkB} ,
	\end{align}
	where $\rho$ and $\bs{J}$ are the charge density and current density, and the elements of $F'^{\mu\nu}$ have been defined analogously to the electromagnetic field as
	\begin{align}
		&\bs{B'}=\nabla \times \bs{A'}\label{def:DarkB} ,\\
		&\bs{E'}=-\nabla A'^0-\frac{\partial \bs{A'}}{\partial t}\label{def:DarkE}.
	\end{align}
	The Maxwell equation for normal photon can be obtained as usual,
	\begin{align}
		&\nabla \cdot \bs{E}=\rho\label{equ:divE} ,\\
		&\nabla\cdot\bs{B}=0\label{equ:divB} ,\\
		&\nabla \times \bs{E}+\frac{\partial \bs{B}}{\partial t}=0\label{equ:rotE} ,\\
		&\nabla \times \bs{B}-\frac{\partial \bs{E}}{\partial t}=\bs{J}\label{equ:rotB}.
	\end{align}
	In addition, one can determine the current in the conductor as,
	\begin{equation}
		\bs{J}=\sigma(\bs{E}-\epsilon\bs{E'})\label{equ:current} ,
	\end{equation}
	and the charge conservation is described as,
	\begin{equation}
		\nabla \cdot \bs{J}+\frac{\partial\rho}{\partial t}=0\label{equ:conservation} .
	\end{equation}
	
	\subsection{ Flat metal plate}
	First, we consider the situation of an infinitely large metal plate with thickness $h$, which interacts with an incoming dark photon field. From Eqs. \eqref{equ:current}, \eqref{equ:conservation} and Maxwell equations for $A$ and $A'$, we have:
	\begin{equation}
		\frac{\partial \rho}{\partial t}+ \sigma\rho + \sigma\epsilon m_{A'}^2A'^{0}=0\label{equ:rho1},
	\end{equation}
	ignoring terms of order $O(\epsilon^2)$.
	From the Maxwell equations and Eq.~\eqref{equ:current} one can derive:
	\begin{equation}
		\nabla^2\bs{E}-\frac{\partial^2 \bs{E}}{\partial t^2}-\sigma\frac{\partial\bs{E}}{\partial t}=\nabla\rho-\sigma\epsilon\frac{\partial\bs{E'}}{\partial t}\label{equ:waveE}.
	\end{equation}
	Similar equation holds for $\bs{E'}$,
	\begin{align}
		\nabla^2\bs{E'}-\frac{\partial^2 \bs{E'}}{\partial t^2}-\sigma\epsilon^2\frac{\partial\bs{E'}}{\partial t}=-\epsilon\nabla\rho-\sigma\epsilon\frac{\partial\bs{E}}{\partial t}\nn\\
		-m_{A'}^2\frac{\partial\bs{A'}}{\partial t}-m_{A'}^2\nabla A'^0\label{equ:waveDarkE} .
	\end{align}
	
	Assume the incoming dark photon takes the form of a plane wave,
	\begin{equation}
		A'^{\mu}(x)=\tilde{A}'^{\mu}e^{ikx}\label{equ:planewave},
	\end{equation}
	where the tilde means the field without plane wave factor and with a parameter $\beta$,
	\begin{equation}
		\beta  \equiv \frac{\sigma}{\sigma-i\omega}\label{def:beta},
	\end{equation}
	the solution to Eq.~\eqref{equ:rho1} is:
	\begin{equation}
		\rho(x)=-\beta\epsilon m_{A'}^2A'^0e^{ikx}\label{sol:rho}.
	\end{equation}
	Therefore, the charge density $\rho$ and current density $\bs{J}$ are both of order $O(\epsilon)$, and from Eq.~\eqref{equ:waveE} the electric field $\bs{E}$ is also of order $O(\epsilon)$. Neglecting terms of order $O(\epsilon^2)$, Eq.~\eqref{equ:waveDarkE} can be simplified to
	\begin{equation}
		\nabla^2\bs{E'}-\frac{\partial^2 \bs{E'}}{\partial t^2}=-m_{A'}^2\frac{\partial\bs{A'}}{\partial t}-m_{A'}^2\nabla A'^0 ,
	\end{equation}
	which is the vacuum equation of $\bs{E'}$. It implies that the dark electric field satisfies the same plane wave function inside the conductor.
	
	Following Eqs.~\eqref{def:DarkE} and \eqref{equ:planewave}, the dark electric field can be written in an explicit form:
	\begin{equation}
		\bs{E'}=(i\omega \tilde{\bs{A}}'-i\bs{k}\tilde{A}'^0)e^{i\bs{k}\cdot\bs{r}-i\omega t}.
	\end{equation}
	Define another parameter $\alpha$
	\begin{equation}
		\alpha \equiv \frac{i\omega\sigma}{m_{A'}^2+i\omega\sigma}\label{def:alpha},
	\end{equation}
	the solution to Eq.~\eqref{equ:waveE} is:
	\begin{equation}
		\bs{E}=\bs{E}_{in}e^{i\bs{k}\cdot\bs{r}-i\omega t},
	\end{equation}
	where
	\begin{equation}
		\bs{E}_{in}=i\epsilon\alpha\omega \tilde{\bs{A}}'-i\epsilon(\beta+\alpha-\beta \alpha) \bs{k} \tilde{A}'^0 \label{sol:Ein}.
	\end{equation}
	
	\begin{figure}
		\centering
		\includegraphics[width=0.8\linewidth]{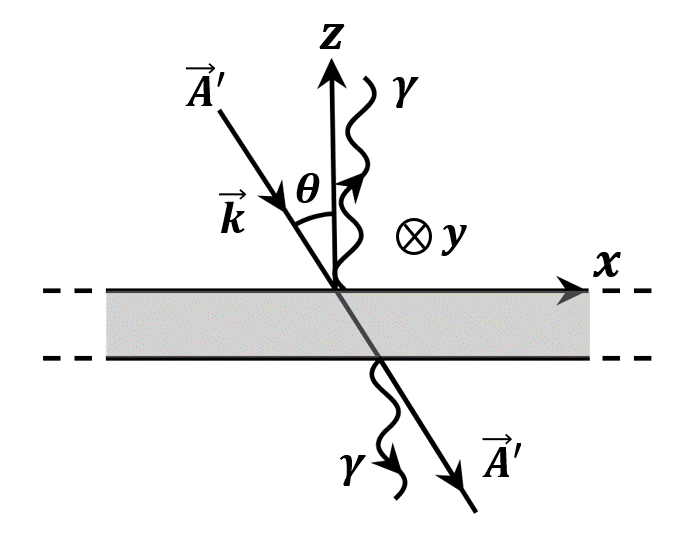}
		\caption{The perfect conductor plate and the DPDM field and EM fields plotted at the order of $\mathcal{\epsilon}$. }
		\label{fig:conductorplate}
	\end{figure}
	
	One can write $\bs{E}_{in}$ more explicitly 
	\begin{align}
		&E_{in,x}=i\epsilon\alpha\omega \tilde{A}'_x-i\epsilon(\beta+\alpha-\beta \alpha)k \tilde{A}'^0\sin\theta ,\\
		&E_{in,y}=i\epsilon\alpha\omega \tilde{A}'_y ,\\
		&E_{in,z}=i\epsilon\alpha\omega \tilde{A}'_z+i\epsilon(\beta+\alpha-\beta \alpha)k \tilde{A}'^0\cos\theta ,
	\end{align}
	using the coordinate system shown in Fig.~\ref{fig:conductorplate}.
	The three equations above describe the electric field inside the conductor, $-h<z<0$. 
	For $z>0$, there is another plane wave going away from the conductor plate,
	\begin{equation}
		\bs{E}=\bs{E}_{up} e^{i(\bs{k}_{up} \cdot \bs{r}_{up} -\omega_{up} t)},
	\end{equation}
	where $\bs{k}_{up}$ and $\omega_{up}$ can be determined by the boundary conditions and the dispersion relation of the photon,
	\begin{align}
		&\omega_{up}=\omega , \\
		&k_{up,x}=k\sin\theta , \\
		&k_{up,z}=\sqrt{\omega^2-k^2\sin^2\theta}.
	\end{align}
	It is worth mentioning that the momentum of dark photons is much smaller than its rest mass,  $k \ll m_{A'}$. Therefore, the wave vector $\bs{k}_{up}$ is almost perpendicular to the conductor plate, with a tiny zenith angle of order $O(k/m_{A'})\sim 10^{-3}$. 
	
	Since the electromagnetic wave in the upper space is an ordinary plane wave solution to the Maxwell equations, the electric field is perpendicular to the wave vector. Together with the boundary conditions, $\bs{E}_{up}$ can be obtained as
	\begin{align}
		&E_{up,x}= E_{in,x} ,\\
		&E_{up,y}= E_{in,y} ,\\
		&E_{up,z}=\frac{i\epsilon k\sin\theta}{\sqrt{\omega^2-k^2\sin^2\theta}}  \nonumber \\
		& \times \left((\beta+\alpha-\beta \alpha)k \tilde{A}'^0\sin\theta-\alpha\omega \tilde{A}'_x \right) .
	\end{align}
	If we ignore the velocity of the dark photon by taking $k\rightarrow 0$, then we have:
	\begin{align}
		&E_{up,x}=i\epsilon\alpha\omega \tilde{A}'_x=\epsilon\alpha E_{x}' ,\\
		&E_{up,y}=i\epsilon\alpha\omega \tilde{A}'_y=\epsilon\alpha E_{y}' ,\\
		&E_{up,z}=0 .
	\end{align}
	The normal magnetic field inside and outside the conductor can be derived accordingly, following the Maxwell equation \eqref{equ:rotE},
	\begin{align}
		&B_{in,x}=i\epsilon\alpha k \tilde{A}'_y\cos\theta ,\\
		&B_{in,y}=-i\epsilon\alpha k(\tilde{A}'_x\cos\theta + \tilde{A}'_z\sin\theta) ,\\
		&B_{in,z}=i\epsilon\alpha k \tilde{A}'_y\sin\theta ,\\
		&B_{up,x}=-i\epsilon\alpha\sqrt{\omega^2-k^2\sin^2\theta} \tilde{A}'_y ,\\
		&B_{up,y}=\frac{i\epsilon \omega}{\sqrt{\omega^2-k^2\sin^2\theta}}[\alpha\omega \tilde{A}'_x-(\beta+\alpha-\beta \alpha)k \tilde{A}'^0\sin\theta] ,\\
		&B_{up,z}=i\epsilon\alpha k \tilde{A}'_y\sin\theta .
	\end{align}
	Following the same procedure, one can obtain the field in the $z<-h$ region. The only difference is an overall phase factor and a minus sign of the z-component of the wave vector. Defining the phase factor as
	\begin{equation}
		\delta \equiv e^{ik h \cos\theta},
	\end{equation}
	the electromagnetic field below the plate can be expressed in terms of its counterpart in the upper space.
	\begin{align}
		&E_{down,x}=\delta \times E_{up,x},\\
		&E_{down,y}=\delta \times  E_{up,y},\\
		&E_{down,z}=-\delta \times  E_{up,z},\\
		&B_{down,x}=-\delta \times  B_{up,x},\\
		&B_{down,y}=-\delta \times  B_{up,y},\\
		&B_{down,z}=\delta B_{up,z}.
	\end{align}
	
	Lastly, the current related to the plate can be obtained, which consists of three parts, the current density $\bs{J}$ and the surface current on the upper surface $\bs{i}_{up}$ and lower surface $\bs{i}_{down}$, respectively. 
	The current density can be obtained by Eq.~\eqref{equ:current},
	\begin{equation}
		\bs{J}=\sigma\epsilon[i\omega(1-\alpha)\bs{A'}+i\bs{k}(1-\alpha)(1-\beta)A'^0].
	\end{equation}
	Considering a perfect conductor, e.g., aluminum, we can have $m_{A'}/\sigma\sim 10^{-9}$, which implies $\alpha \simeq 1$.  
	Therefore, the current density vanishes inside the conductor. It also implies $\bs{E}$ is at the order of $\mathcal{O}(\epsilon)$ and cancels with the dark electric field as, 
	\begin{align}
		\bs{E}_{in} - \epsilon \bs{E}_{in}' =0 .
	\end{align}
	
	For surface current, one can obtain their value on both sides of the metal plate using the boundary condition of the magnetic field,
	\begin{align}
		&i_{up,x}=B_{in,y}-B_{up,y} ,\\
		&i_{up,y}=B_{up,x}-B_{in,x} ,\\
		&i_{down,x}=-\delta \left(B_{up,y}+B_{in,y} \right)  ,\\
		&i_{down,y}=\delta \left(B_{in,x}+B_{up,x}\right) .
	\end{align}
	In summary, under the non-relativistic condition, good conductor, and thin plate assumptions, we arrive at the simplified results for the currents,
	\begin{align}
		&i_{tot,x}=i_{up,x}+i_{down,x} \approx -2i\epsilon m_{A'}A'_x , \\
		&i_{tot,y}=i_{up,y}+i_{down,y} \approx -2i\epsilon m_{A'}A'_y ,\\
		&\bs{J}=0 .
	\end{align}
	
	Therefore, the flat metal plate behaves like an oscillating electric dipole $\bs{p}$, and its time derivative is
	\begin{equation}
		\dot{\bs{p}}=-2i\epsilon m_{A'} \bs{A'_{\tau}} \Delta S, 
	\end{equation}
	where $\bs{A'_{\tau}}$ represents the projection of $\bs{A'}$ to the metal plane, and $\Delta S$  is the related surface area unit. We introduce two vectors $\bs{n}$ and $\bs{n_0}$, representing the normal direction of the metal plate and the direction of $\bs{A'}$ respectively. Then the projection direction $\bs{\tau}$ can be defined as,
	\begin{equation}
		\bs{\tau} \equiv \bs{n_0}-(\bs{n_0}\cdot\bs{n})\bs{n},
	\end{equation}
	and $\bs{A'_{\tau}} = \bs{\tau} |\bs{A'}|$.
	\\

	\subsection{DPDM-Induced EM flux}
	For the simulation of the response of the dish antenna to DPDM, the dish surface is divided into small pieces. The size of each piece is much smaller than the wavelength of the induced EM signal $\lambda$, but much larger than the thickness of the metal plate. 
	Exploiting the results of dipole radiation, we obtain the induced EM field at position $\bs{r}$,
	\begin{align}
		&\bs{E}=-\frac{\epsilon m_{A'}^2 |\bs{A'}|}{2\pi}\nn  \\
		&\int[\bs{\tau}\times(\bs{r}-\bs{r'})]\times(\bs{r}-\bs{r'})
		\frac{e^{im_{A'}|\bs{r}-\bs{r'}|+i\bs{k}\cdot\bs{r'}}}{|\bs{r}-\bs{r'}|^3}\d S'\label{equ:E} ,
	\end{align}
	\begin{equation}
		\bs{B}=-\frac{\epsilon m_{A'}^2 |\bs{A'}|}{2\pi}\int\bs{\tau}\times(\bs{r}-\bs{r'})\frac{e^{im_{A'}|\bs{r}-\bs{r'}|+i\bs{k}\cdot\bs{r'}}}{|\bs{r}-\bs{r'}|^2}\d S'\label{equ:B} .
	\end{equation}
	It should be mentioned that for the FAST telescope, the DM velocity has a non-trivial contribution to the phase factor in the above equations because the telescope size is comparable to the wavelength of DPDM $\lambda'$. 
	Therefore, the time-averaged energy flux density of the induced EM wave can be written as
	\begin{equation}
		\langle \bs{S'}\rangle =\frac{1}{2}\mathrm{Re}(\bs{E}\times\bs{B^*})\label{equ:S}.
	\end{equation}
	The value of $\bs{A'}$ can be calculated from the local DM energy density. From the Lagrangian~\eqref{equ:lagrangian}, we obtain
	\begin{equation}
		\rho_{\rm DM}=\frac{1}{2}m_{A'}^2|\bs{A'}|^2\label{equ:rho}.
	\end{equation}
	With Eqs.~\eqref{equ:E}, \eqref{equ:B}, \eqref{equ:S}, and \eqref{equ:rho}, the energy flux density can be calculated numerically.

	\begin{figure}
		\centering
		\includegraphics[width=0.7\linewidth]{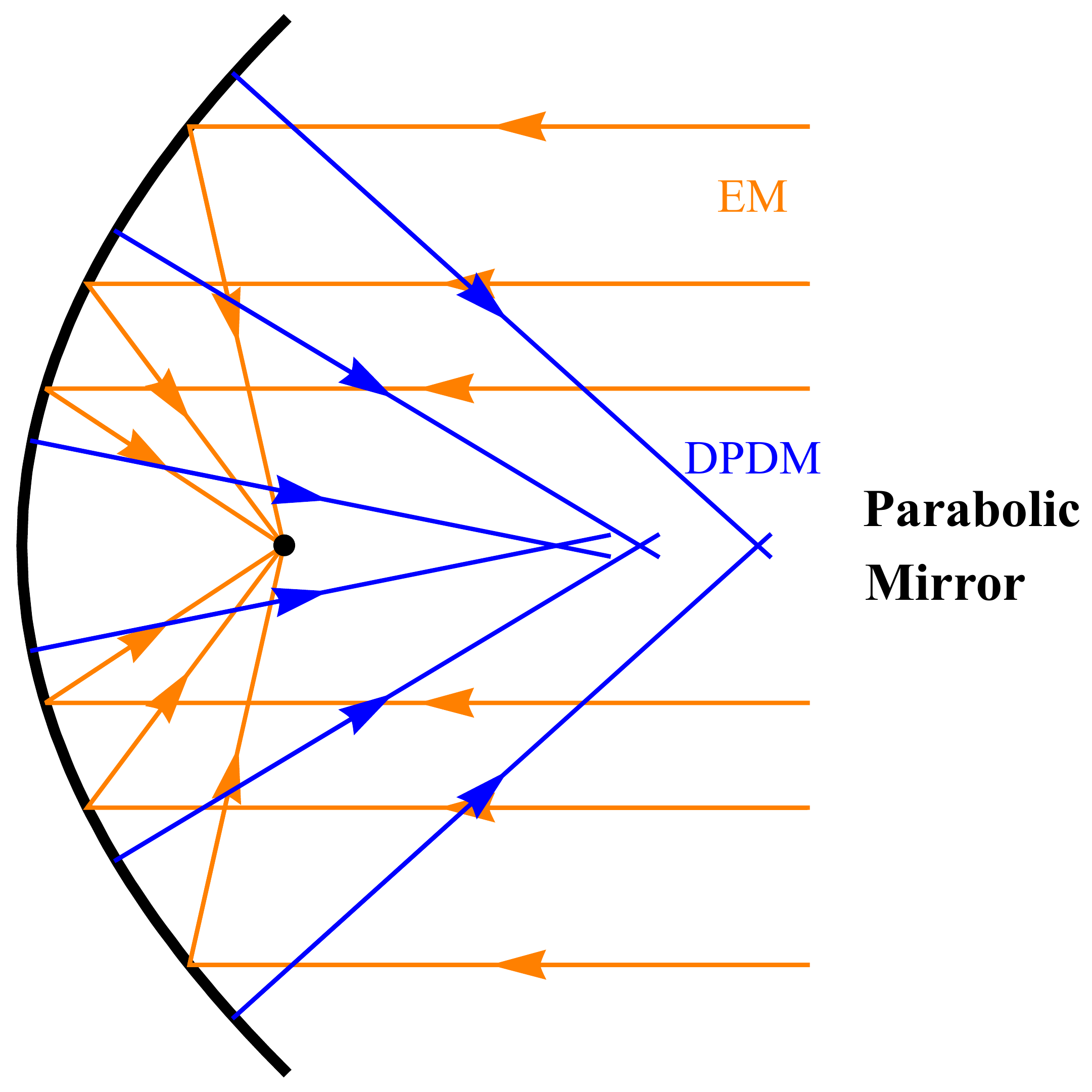}
		\includegraphics[width=0.7\linewidth]{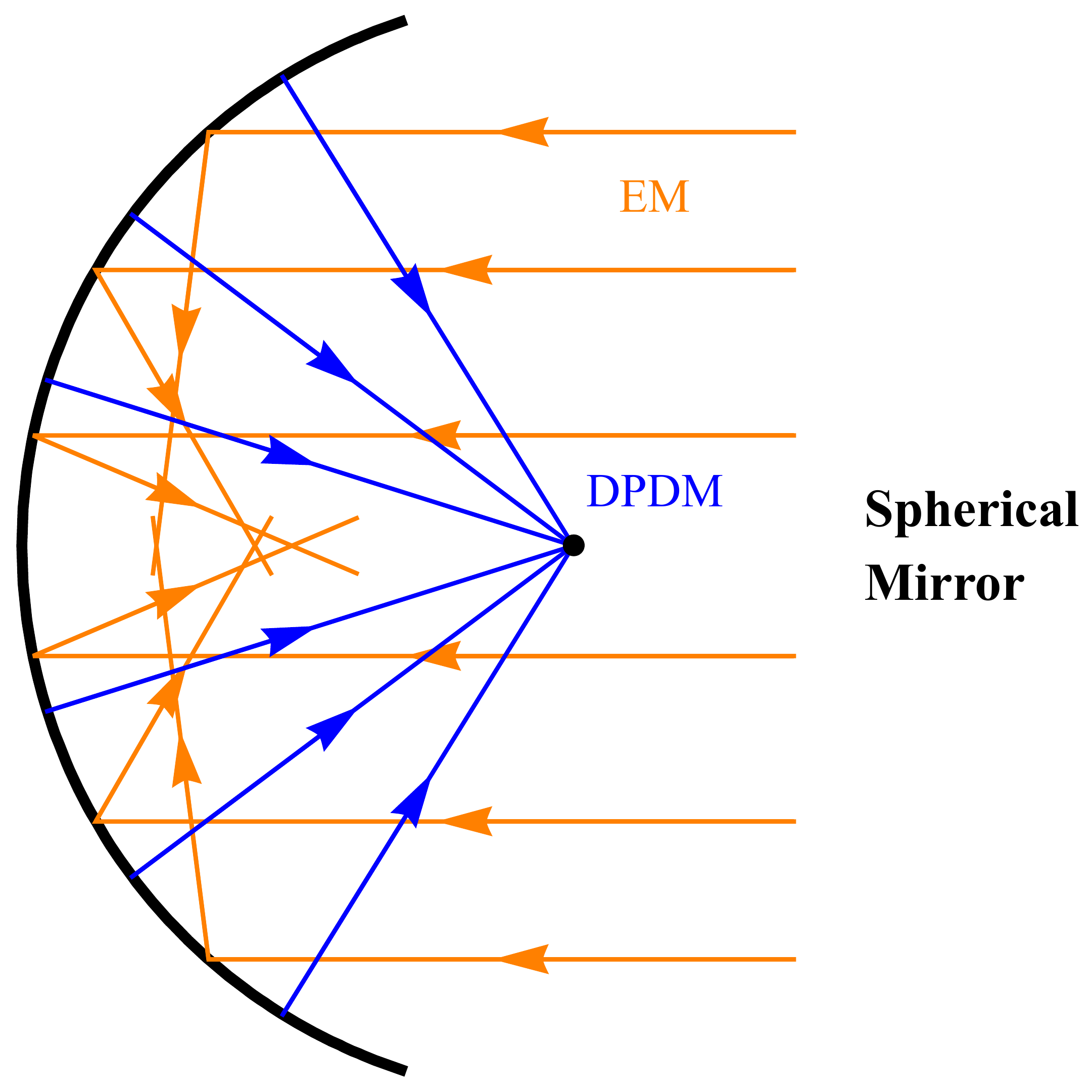}
		\caption{
			The ray-optics schematic plots show the reflected EM flux induced by normal EM waves (orange) and DPDM waves (blue) in the parabolic mirror (top) and spherical mirror (bottom) setups. In the parabolic mirror setup, parallel EM waves are focused on the focal point, while the DPDM-induced EM waves are spread into a focal volume. In the spherical mirror setup, the opposite occurs, with the DPDM-induced EM waves being focused at the spherical center.
		}
		\label{fig:two-mirrors}
	\end{figure}
	
	For example, we can calculate the energy flux going into the feed analytically for a spherical metal plate with a radius $R$ where the feed is placed at the center.
	Assuming the size of the antenna is much smaller than the wavelength of dark photons, the energy flux density at the center is
	\begin{align}\label{eq:S_over_rho}
		& \frac{|\bs{S'}|}{\rho_{\rm DM}} =\frac{1}{3}\pi^2\epsilon^2\frac{R^2}{\lambda^2}s_\gamma s^2_{\theta_0} \times \\ &\sqrt{\left(2-3 c_{\theta_0}+c^3_{\theta_0}\right)^2 c^2_\gamma+4\left(1-c_{\theta_0}^3 \right)^2 s^2_\gamma} \nonumber ,
	\end{align}
	where $s_\gamma = \sin \gamma$, $c_{\theta_0} = \cos\theta_0$. The angle $\theta_0$ describes how large is the spherical surface with $\theta_0= 0 $ for the surface shrinking to a point and $\theta_0= \pi $ for the surface becoming a full sphere.
	The $\gamma$ describes the angles between the polarization vector of $\bs{A}'$, $\bs{n}_0$, and the z-direction.
	The large ratio $R^2/\lambda^2$ explicitly shows the enhancement from the ray optics focusing effect because the induced EM wave is perpendicular to the surface and is greatly enhanced for a spherical surface. For $A'$ polarization in the z-direction, $\gamma = 0 ~{\rm or}~\pi$, the energy flux density will go to zero. For other $\gamma$, one can calculate the best $\theta_0$, which has the largest energy flux density. This analytic formula helps people design the spherical mirror, which can help compromise efficiency and cost. 
	
	For the FAST telescope, the part of the dish that faces the feed deforms into a 300-meter diameter parabolic shape with the feed located at the focus. 
	As a result, for the DPDM signal, we no longer have the enhancement factor $R^2/\lambda^2$.

	At the end of this subsection, we demonstrate the difference between the parabolic mirror and spherical mirror by Fig.~\ref{fig:two-mirrors} drawing in the ray-optics schematic plots. The induced EM flux from parallel incident EM wave (orange) and DPDM (blue) are shown, where we assume the coming direction of the incident EM wave matches the parabolic mirror. At the same time, we do not need to specify the incoming direction of DPDM, because the induced EM waves are always perpendicular to the mirror's surface.  
	For the parabolic mirror, the parallel EM waves are focused in the focal point, while the DPDM-induced EM waves are spread into a focal volume. For the spherical mirror, the vice versa happens where DPDM induced EM wave focused in the spherical center. Therefore, there is an enhancement for parallel EM wave incident on the parabolic mirror, while for DPDM incident on the spherical mirror.

	\subsection{Calculation of the ${\cal C_{\rm dish}}$ parameter for dish telescopes}
	\label{sec:C-dish-numeric}
	
	Here we present the detailed calculation of the ${\cal C_{\rm dish}}$ parameter, used in estimating the future reaches of the FAST and SKA1-Mid telescopes. The ${\cal C_{\rm dish}}$ parameter is defined in Eq.~(\ref{eq:dish}) for dish telescopes to calculate the projections of their sensitivities. In Eq.~\eqref{eq:dish}, $I^{\rm eqv}_{\rm dish}$ is defined as the equivalent flux density of a plane EM wave from infinite faraway that will produce the same signal strength as the DPDM. Therefore, to calculate ${\cal C_{\rm dish}}$, we also need to simulate the EM flux density at the position of the feed induced by the plane EM wave from an infinite faraway. The simulation is parallel to the DPDM signal simulation discussed above. The only difference is that the phase of the dipole oscillations in the metal plate of the reflector is now controlled by the plane EM wave, such that the reflected EM wave will focus on the feed position. 
	
\begin{figure}
		\centering
		\includegraphics[width=0.95\linewidth]{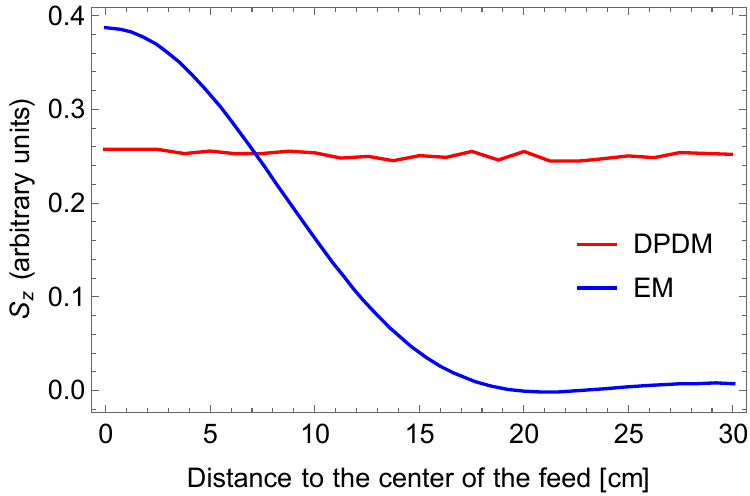}
		\caption{The distributions of $S_z$ at the feed induced by EM (blue) and DPDM (red) fields at $1$ GHz. The diameter of the feed is about 20 cm.}
		\label{fig:Sz}
	\end{figure}

	The feed size is usually designed to be around the wavelength $\lambda$ to optimize the signal. Therefore, to estimate the sensitivity, we assume the feed of the future detectors to be a round shape with a diameter equal to $\lambda$. And both the DPDM-induced flux and EM plane wave-induced flux received by the feed are calculated as 
	\bea\label{eq:F}
	{\cal F} = \int_{r<\lambda/2} r dr d\theta \langle S_z \rangle \ , 
	\eea
	where $S_z$ is the flux along the direction of the receiver of the feed. As an illustration, the distributions of $S_z$ for both the EM and DPDM-induced signals at the feed are shown in Fig.~\ref{fig:Sz} for the FAST  telescope. We can see that $S_z$ for the EM-induced signal drops quickly when it is away from the center of the feed. However, $S_z$ for the DPDM-induced signal is almost flat. Then the explicit expression for ${\cal C_{\rm dish}}$ can be written as
	\bea\label{eq:CC}
	{\cal C_{\rm dish}} = \frac{{\cal F}_{\rm DPDM}[\rho_0]}{{\cal F}_{\rm EM}[\rho_0]} \times \frac{1}{\epsilon^2} \times \frac{{\cal A}}{\lambda^2}  \ ,
	\eea
	where $\cal A$ is the area parameter. For ${\cal F}_{\rm DPDM}$, $\rho_0$ is the energy density of the DPDM, and for ${\cal F}_{\rm EM}$, $\rho_0$ is the energy density of the parallel EM field. Since both the ${\cal F}_{\rm DPDM}$ and ${\cal F}_{\rm EM}$ are linear in $\rho_0$, the dependence on $\rho_0$ is canceled, and therefore ${\cal C}_{\rm dish}$ is independent of $\rho_0$. 
	
	\begin{figure}
		\centering
		\includegraphics[width=0.95\linewidth]{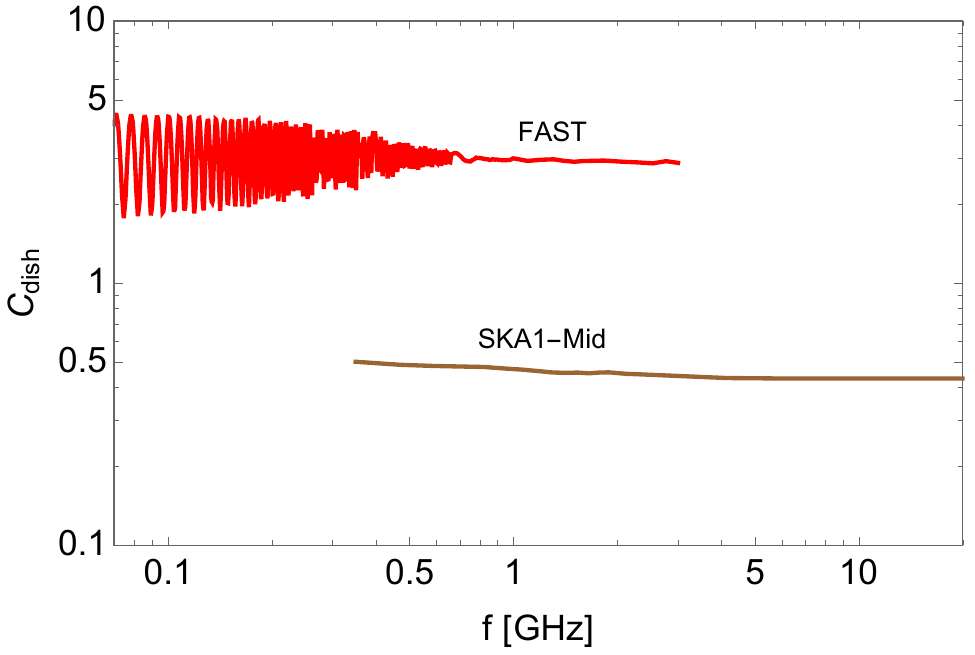}
		\caption{The values of ${\cal C}_{\rm dish}$ for FAST and SKA1-Mid telescopes respectively.}
		\label{fig:Cdish}
	\end{figure}
	
	The values of ${\cal C_{\rm dish}}$ for calculating the projective sensitivities of the FAST and SKA1-Mid telescopes are shown in Fig.~\ref{fig:Cdish}. One can see that the values of ${\cal C_{\rm dish}}$ are independent of $f$ in the high-frequency region. However, in the low-frequency region, when $\lambda'$ becomes larger than the size of the telescope, ${\cal C_{\rm dish}}$ starts to show an oscillation pattern as shown by the red curve in Fig.~\ref{fig:Cdish}. The reason is that when the wavelength of DPDM, $\lambda'$, is significantly larger than the size of the telescope, the induced EM waves on the reflector plate share the same phase and form a stationary wave-like structure inside the reflector bowl. As a result, the positions of the nodes and antinodes change with the frequency, which causes the oscillation pattern of $C_{\rm dish}$ for FAST in the low frequency region.

	\subsection{The calculation of $I_{\rm FAST}^{\rm eqv}$}
	\label{sec:C-FAST-simulation}

	\begin{figure}
		\centering
		\includegraphics[width=0.95\linewidth]{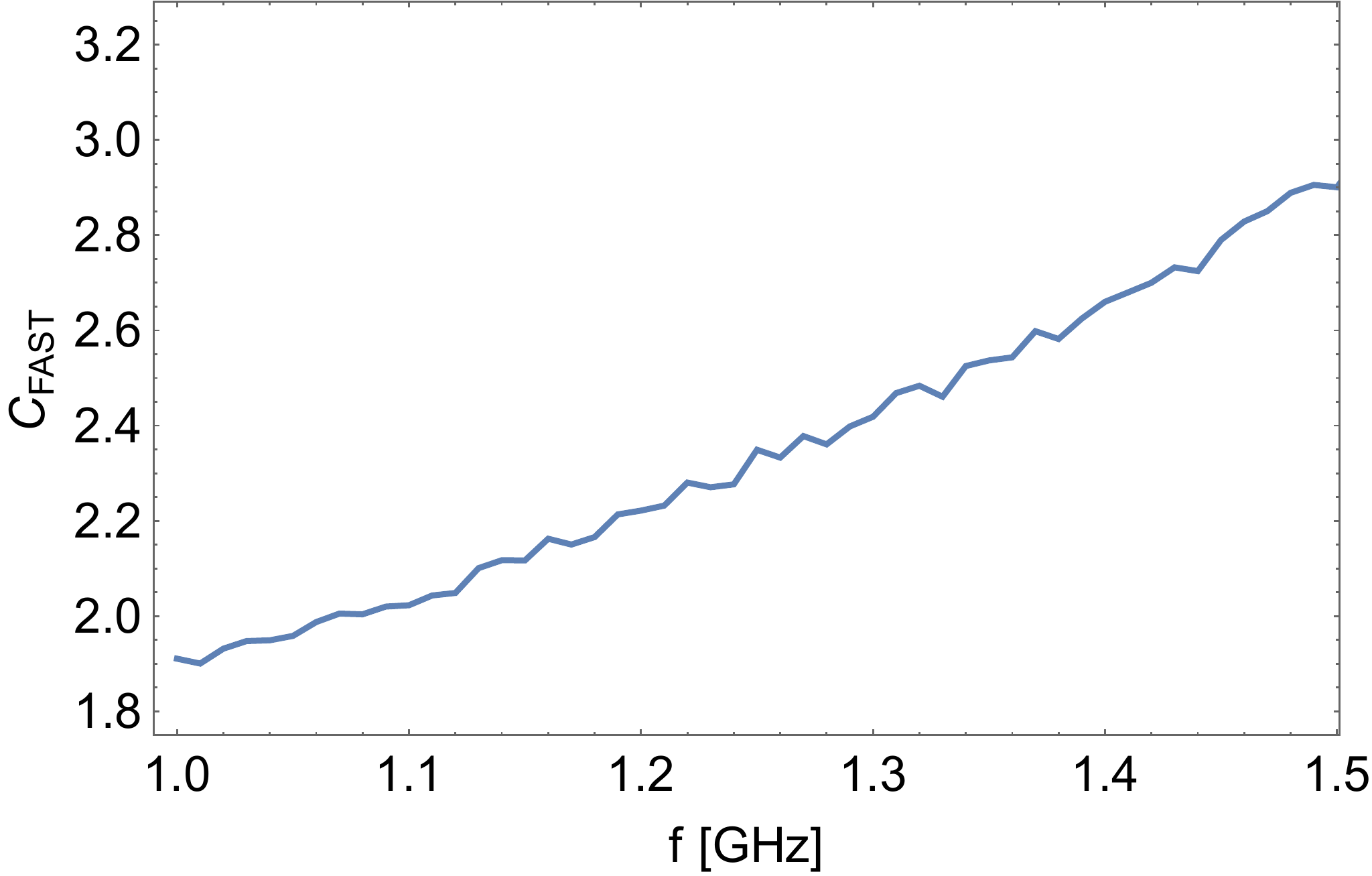}
		\caption{The values of ${\cal C}$ for FAST real data analysis in the data range $1-1.5$ GHz.}
		\label{fig:CFAST}
	\end{figure}
	
	The data we acquire from the FAST observation has been interpreted as the plane EM wave from distant stars. Therefore, to calculate the constraint, we first need to calculate $I_{\rm FAST}^{\rm eqv}$. The current feed of the FAST telescope is composed of 19 beam receivers. The diameter for each beam receiver is $D_b = 20$ cm. Therefore, when calculating ${\cal F}$ in Eq.~\eqref{eq:F}, the upper limit of the radial integral needs to be replaced by $D_b/2$, and the value of ${\cal C}$, in this case, depends on the frequency.
	We have simulated the values of ${\cal C}$ at different frequencies, averaging over the dark photon velocity distribution and polarizations.
	The result is shown in Fig.~\ref{fig:CFAST}. Then, the equivalent EM flux density induced by DPDM can be expressed as
	\bea\label{eq:I_FAST_eqv}
	I_{\rm FAST}^{\rm eqv} = \epsilon^2 \rho_{\rm DM} \times \frac{\lambda^2}{{\cal A}_{\rm FAST}} \times{\cal C}_{\rm FAST}(f) \ .
	\eea
	For example, using the simulated ${\cal C}_{\rm FAST}(f=1~{\rm GHz})=1.91$, we have $I_{\rm FAST}^{\rm eqv} \approx 0.035 \epsilon^2 ~{\rm W/m^2}$ at $1$~GHz with the area parameter ${\cal A}_{\rm FAST}=\pi(150~{\rm m})^2$ plugged in. 
	Later, we will compare the simulation result with the observation data of FAST and set limits on the kinetic mixing coupling $\epsilon$.

	\begin{figure}
		\centering
		\includegraphics[width=0.8\linewidth]{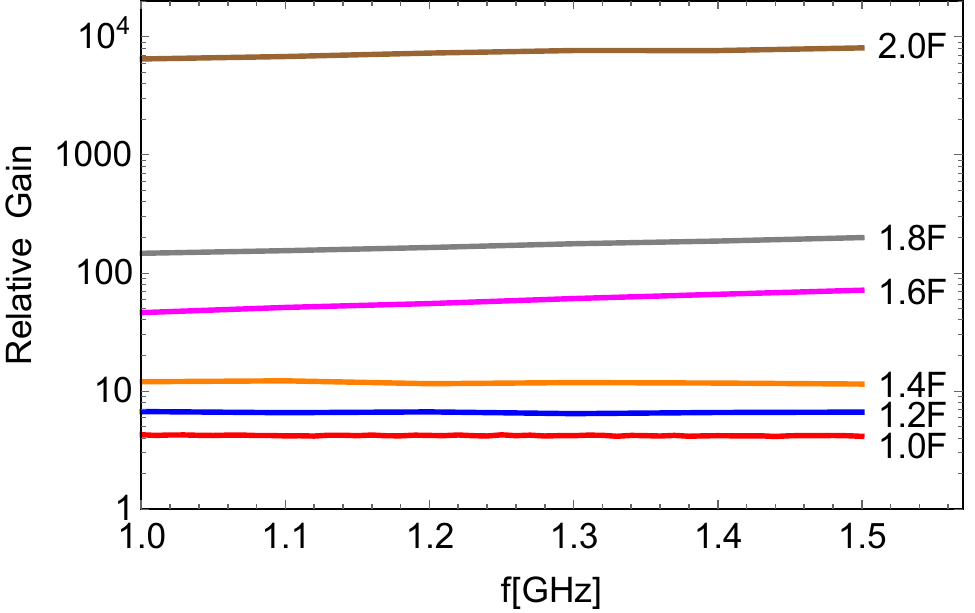}
		\caption{The relative gain provided by the FAST dish in comparison to the direct signal, with different colors representing different heights of the feed above the dish. The height of the focal surface, $F$, is 140 meters, and the feed is currently located at $1F$.}
		\label{fig:relative_gain}
	\end{figure}
	
	The detector in the feed can also directly detect a signal generated by DPDM oscillation. However, accurately calculating this signal is difficult due to the complexity of the feed's structure. To estimate the direct signal, we use the equivalent electric field method developed in this work. The equivalent electric field at the feed's position can be expressed as:
	\bea
	\bE^{\rm eqv}_{\rm direct} = \epsilon E_0' \ ,
	\eea
	where the spatial phase is omitted since we have only one {\it antenna}. Then, we can estimate the equivalent flux detected by the feed as
	\bea
	I^{\rm eqv}_{\rm direct} = \frac{2}{3} \epsilon^2 {E_0'}^2 = \frac{2}{3} \epsilon^2 \rho_{\rm DM}\ ,
	\eea
	where the factor of $2/3$ accounts for the fact that the detector is designed to detect transverse polarized EM waves. By using the ratio $I^{\rm eqv} / I^{\rm eqv}_{\rm direct} $, we can estimate the relative gain provided by the dish to the direct signal. This ratio is shown as a function of frequency in Fig.~\ref{fig:relative_gain}, where it can be seen that the relative gain is almost frequency-independent in the sensitive region of FAST. The different colors in the figure correspond to different heights of the feed. In the current observation, the feed is located at $1F \approx 150$ meters, and the relative gain is about four. This implies that the reflector-induced signal is about four times larger than the direct signal, which can be understood from the geometric picture shown in Fig.~\ref{fig:two-mirrors}. Since the feed is located at the half-radius position of the focal surface, and the reflector-induced EM wave is perpendicular to the reflector surface, the energy flux of the reflector-induced EM wave at the focal surface is enhanced by roughly a factor of four due to energy conservation. This explains the four-fold relative gain between the dish-induced signal and the direct signal. 
	
	Furthermore, the sensitivity of FAST could be significantly improved by raising the feed to higher locations. The calculations indicate that by placing it at a distance of $2F\approx 300$ meters, the relative gain could increase by a factor of $6000$, as demonstrated in Fig.~\ref{fig:relative_gain}. Thus, it would be worthwhile to cooperate with the FAST team to determine if raising the feed is feasible within their mechanical system.

	One may be concerned about interference between the direct and dish-induced signals. However, the feed is positioned approximately 140 meters above the reflector, resulting in significant suppression of the interference effect. Using Eq.~(\ref{eq:Smn}), we can estimate the interference suppression factor to be approximately 0.3. Additionally, the size of the reflector will induce cancellations, further reducing the interference contribution. As a result, the interference contribution can be neglected compared to the direct signal.
	Furthermore, due to the complex nature of the feed structure, accurately simulating the direct and interference contributions is difficult. 
	Therefore, in this study, we only use the reflector-induced signal to calculate the FAST constraint.
	In addition, the interference between reflector and feed, along with the direct signal may result in an $\mathcal{O}(10\%)$ uncertainty for the FAST limits.

	\section{Response calculation of the dipole antenna}
	\label{sec:C-for-dipole}
	
	The LOFAR and SKA1-Low telescopes are composed of dipole antennas. The antennas are composed of metal bars with different lengths to achieve wideband observations. For example, the antennas of LOFAR LBA have two uneven dipole components, whereas SKA1-Low uses log-periodic dual-polarized antennas. Therefore, accurate calculations of the responses require complex antenna configurations, which is hard to get analytic results and physics intuition. One needs to do accurate numerical simulations of the fields and currents inside the antennas for both DPDM-induced and electromagnetic wave-induced signals. We will leave this for future studies when applying to the LOFAR and SKA1-Low observational data. 
	
	Instead, we take the commonly used linear dipole antenna as an example and work out the response factor ${\cal C}_{\rm dipole}$ for DPDM analytically. In general, the induced electric currents on the wire should be calculated, which has been done for distant parallel EM waves in textbook \cite{chatterjee1996antenna}. We will show that the result for DPDM can be obtained smartly by comparing it to an incoming EM wave with $\theta = \pi/2$. We prove that ${\cal C}_{\rm dipole} \geq 1$ is exact for DPDM signals and explain its physics.
	As a result, we use ${\cal C}_{\rm dipole} = 1$ to estimate the projections of the sensitivities for LOFAR and SKA1-Low. These sensitivities should be considered as \textit{conservative} estimates.

	\begin{figure}
		\centering
		\includegraphics[width=0.8\linewidth]{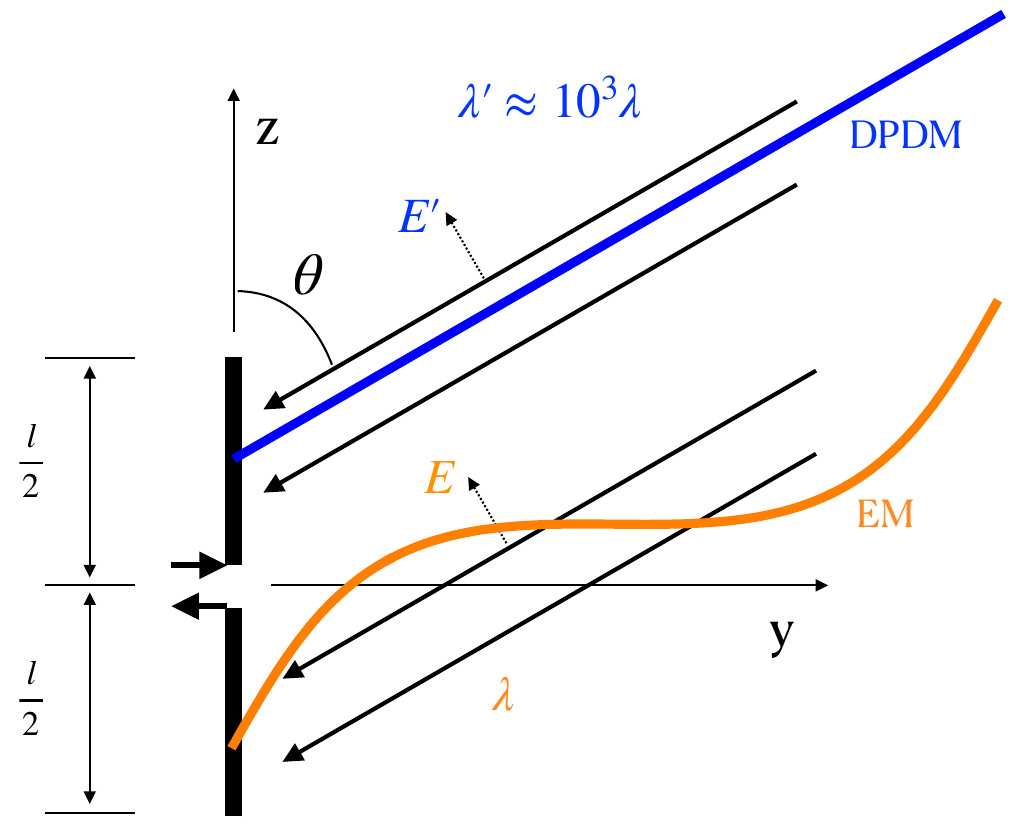}
		\caption{The illustration of a linear dipole antenna with a length $l$. The DPDM and EM waves propagate with an angle $\theta$ to the z-axis. Their polarizations are assumed to be the same. The DPDM wavelength $\lambda'$ is about 1000 times longer than the EM wavelength. Thus, the $E'$ field variation can not be seen in the figure.}
		\label{fig:antenna}
	\end{figure}
	
	The picture of an equal arm linear dipole antenna is shown in Fig.~\ref{fig:antenna}, where the wire is placed along the z-direction. 
	The DPDM and EM waves propagate with an angle $\theta$ to the z-axis. Moreover, their polarizations are assumed to be the same. Since the DPDM wavelength $\lambda' \approx 10^3 \lambda$, and given the fact that the length of linear antenna $ l \sim \mathcal{O}(1) \lambda$ in the antenna design, the $E'$ field variation is hard to see in the figure. Without loss of generality, the electric fields $\bs E$ and ${\bs E'}$ are set to be in the y-z plane because the component in the x direction can not generate an electric current in the linear antenna. 
	
	For incoming EM and DPDM waves in Fig.~\ref{fig:antenna}, their electric fields on the wire can be described as,
	\begin{align}
		{\bs E}(t,z, \theta)  & = {\bs E}_0 e^{i \left( \omega t  + k \cos{\theta} z + \phi_0\right)} , \\
		{\bs E}^{'}(t,z,\theta  )  & = {\bs E}^{'}_0 e^{i \left( \omega t  + k' \cos{\theta} z + \phi'_0\right)}  \approx  {\bs E}^{'}_0 e^{i \left( \omega t   + \phi'_0\right)}, \label{eq:Eprime}
	\end{align} 
	where we have taken the wire at $y = y_0 = 0$, while $\phi_0$ and $\phi'_0$ are their associate phases.
	The oscillation frequency is the same, $\omega = k c$, but the relation $k' \approx 10^{-3} k$ relates the wave number $k$ and $k'$.
	In the second equality of Eq.~\ref{eq:Eprime}, we have used $k' z \ll 1$ according to the fact that $ l \sim \mathcal{O}(1) \lambda$, with the EM wavelength $\lambda = (2\pi)/k$.

	First, we study the special case of the wave that comes from the y-direction, $\theta = \pi/2$, which serves as a benchmark for calculation. In this case, we have $\cos \theta = 0$. Therefore the electric fields for EM and DPDM are the same. Recall their coupling to currents, $e j^\mu_{\rm em} \left(A_\mu  - \epsilon   A'_\mu \right)$, it is clear that $\mathcal{C}_{\rm dipole} = 1$ in this case.
	
	Next, we consider general $\theta$ incidence, which has an \textit{important observation} that the z-component of ${\bs E}^{'}$ is
	\begin{align}
		{\bs E}^{'}_z(t,z,\theta  )  =\sin\theta \frac{|{\bs E}^{'}_0 |}{|{\bs E}_0 |} \times {\bs E}(t,z, \pi/2).
	\end{align}
	It means that the effect of a DPDM wave from incident angle $\theta$ is the same as an EM wave incident from $\theta = \pi/2$, with a suppression factor $\sin\theta $, because only the ${\bs E}^{'}_z$ component can induce a current in the linear wire.
	
\begin{figure}
		\centering
		\includegraphics[width=0.95\linewidth]{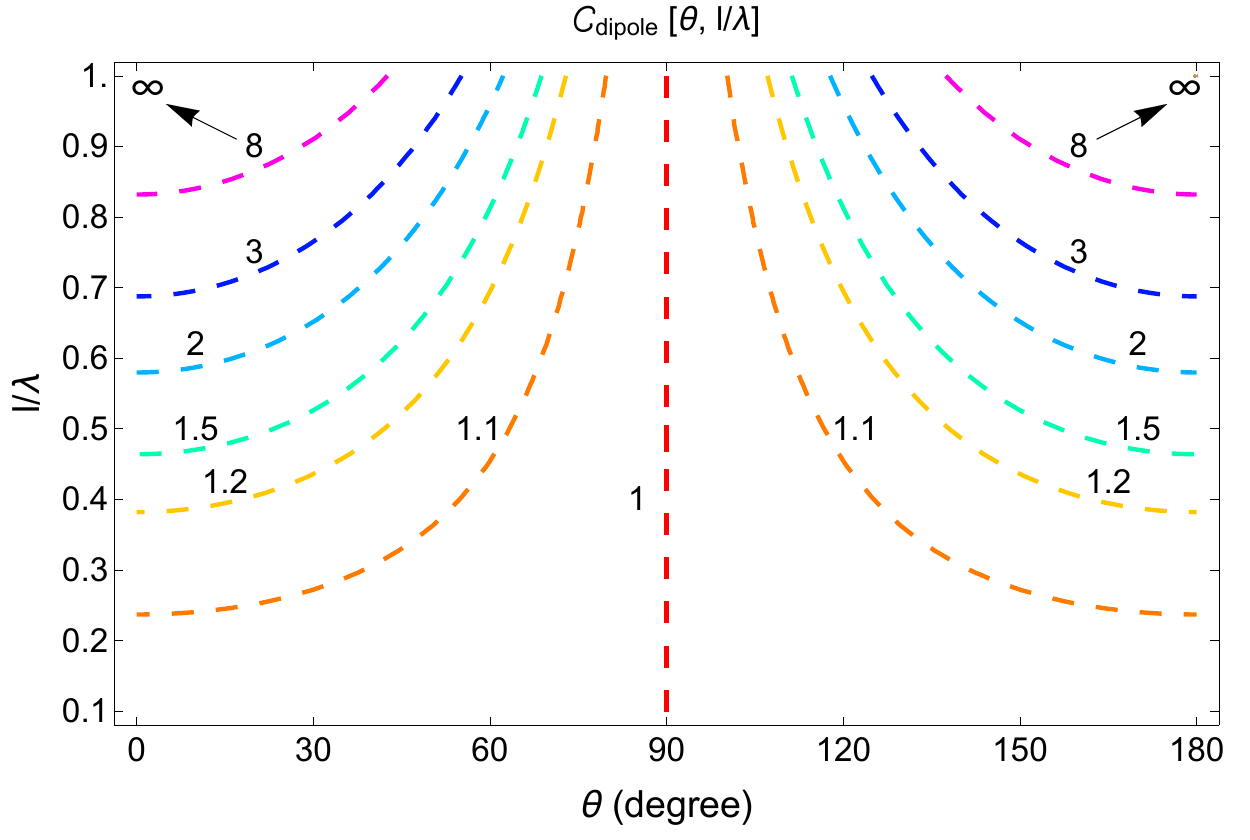}
		\caption{The response factor $\mathcal{C}_{\rm dipole}$ for DPDM incident into a linear dipole antenna, as a function of incoming angle $\theta$ and wavelength $\lambda$. The ratio $l/\lambda = 0.5$ and $1$ represent the half-wavelength and full-wavelength dipole antennas, respectively. An exact relation $\mathcal{C}_{\rm dipole} \geq 1$ holds.}
		\label{fig:Ctheta}
	\end{figure}

	Therefore, to compare DPDM and EM waves both from $\theta$ angle, we need to compare the effects from EM waves for $\theta \neq \pi/2$ and $\theta = \pi/2$. Fortunately, this is described by the far-field radiation pattern function \cite{chatterjee1996antenna} as
	\begin{align}
		f_{\rm EM}(\theta) = \frac{\cos\left(\frac{k l }{2} \cos\theta \right) - \cos\left( \frac{k l }{2} \right) }{\sin\theta}.
	\end{align}
	As a result, the response factor $\mathcal{C}_{\rm dipole}$ for DPDM incident from any $\theta$ is given by,
	\begin{align}
		\mathcal{C}_{\rm dipole}[\theta, l/\lambda] =\left( \sin\theta \times \frac{f_{\rm EM}(\pi/2)}{f_{\rm EM}(\theta)}\right)^2 .
	\end{align}
	In Fig.~\ref{fig:Ctheta}, we plot contours for $\mathcal{C}_{\rm dipole}$, for $\theta \in [0, \ang{180}]$ and $l/\lambda \in [0.1,1]$. The commonly used linear dipole antennas are the half-wavelength dipole antenna with $l/\lambda =0.5$ and $l/\lambda =1$ for the full-wavelength dipole antenna. For $l/\lambda > 1$, the currents in the rod start to cancel each other. Thus, we focus on the region $l/\lambda \leq 1$.

	In Fig.~\ref{fig:Ctheta}, the results show that $\mathcal{C}_{\rm dipole} \geq 1$ is exact for all directions and wavelengths. For $\theta = \pi/2$, it is precisely equal to 1, as explained. It blows up in the corner of $\theta =0, \pi$ and $l/\lambda =1$ because the full-wavelength dipole antenna is usually more directional. 
	
	The reason that the DPDM signal has $\mathcal{C}_{\rm dipole} \geq 1$ feature can be understood in the following. For DPDM, since $\lambda' \gg l$, the antenna wire always feels a uniform {\it dark} electric field. Therefore, the induced current is only weakened by the polarization projection. For the EM case, however, the antenna wire feels the spatial oscillation of the electric field, such that the electric field cancels with each other when driving the electric current in the wire. As a result, the EM-induced electric current is weaker than the DPDM signal.   
	The cancellation argument can also work for complex experiment setups. Thus we expect that taking $\mathcal{C}_{\rm dipole} = 1$ is \textit{conservative}.

	\section{Correlation of the antennas}
	\label{sec:correlation-of-antenna}
	
	The electric field induced by DPDM at each location can be expressed as,
	\begin{align}
		E\left(\vec{x}, ~t \right) = \int_{< k_{\rm esc}} \frac{d^3 k}{(2\pi)^3} b e^{-\frac{k^2}{ k_0^2} } \times  E_0 e^{i \left( \vec{k}\cdot \vec{x} - \omega t  + \theta(\vec{k})\right)},
		\label{eq:EfieldinSHM}
	\end{align}
	where $\theta(\vec{k})$ is a random phase associated with the momentum $\vec{k}$ and the energy is $\omega^2  = \vec{k}^2 + m_{A'}^2$. We average the plane wave function using the truncated Maxwellian distribution (known as the Standard Halo Model), where $b$ is a normalization factor. We have $k_0 = m_{A'} v_0$ and $k_{\rm esc} = m_{A'} v_{\rm esc}$, where $v_0 \approx 235 ~{\rm km/s}$ is most probable velocity and $v_{\rm esc}$ is the escape velocity \cite{McMillan:2009yr, Bovy:2009dr}.
	
	Next, we calculate the correlation of the field at the same location, which should return to its dark photon energy density,
	\begin{align}
		& \rho_{\rm DM}  = \left \langle E\left(\vec{x}, ~t \right) E^*\left(\vec{x}, ~t \right) \right \rangle , \\
		& = \int_{< k_{\rm esc}} \frac{d^3 k}{(2\pi)^3}\int_{< k_{\rm esc}} \frac{d^3 k'}{(2\pi)^3} b^2 \left| E_0 \right|^2 e^{-\frac{k^2 + {k'}^2}{k_0^2}} e^{i \left( \vec{k} - \vec{k'} \right)\cdot \vec{x} }, \nonumber \\
		& \times e^{- i \left(  \omega -\omega' \right)t}  \left \langle e^{ i \left( \theta(\vec{k}) - \theta(\vec{k'}) \right)} \right \rangle .
	\end{align}
	Due to the randomness, we assume there is no correlation between different phases, thus 
	\begin{align}
		\left \langle e^{ i \left( \theta(\vec{k}) - \theta(\vec{k'}) \right)} \right \rangle = a (2 \pi)^3 \delta^3 \left( \vec{k} - \vec{k'} \right),
	\end{align}
	where $a$ is a dimensionful constant. Therefore, we can explicitly work out the energy density of dark photons as
	\begin{align}
		\rho_{\rm DM} & = \frac{a b^2}{32 \pi^{2}} k_0^3 |E_0|^2 \left( \sqrt{2 \pi} {\rm erf}(\sqrt{2}z )- 4 z e^{-2 z^2} \right), \\
		&	\xrightarrow{v_{\rm esc} \to \infty} \frac{a b^2}{8 (2 \pi)^{3/2}} k_0^3 |E_0|^2,
		\nonumber 
	\end{align}
	with $z \equiv k_{\rm esc }/k_0$ and ${\rm erf}$ is the error function. In the second line, we take the limit of large $v_{\rm esc}$. The result shows that the energy density is uniform in time and space.
	
	Similarly, we calculate the correlation of signals at different locations. The full formula using Standard Halo Model is given below,
	\begin{align}
		&	\mathcal{S}(\Delta \vec{x})  \equiv \frac{\left \langle E\left(\vec{x}_1, ~t \right) E^*\left(\vec{x}_2, ~t \right) \right \rangle}{\sqrt{\left \langle E\left(\vec{x}_1, ~t \right) E^*\left(\vec{x}_1, ~t \right) \right \rangle \left \langle E\left(\vec{x}_2, ~t \right) E^*\left(\vec{x}_2, ~t \right) \right \rangle }} , \nonumber \\
		& = \frac{e^{- i  y z}}{\sqrt{2} y \left( \sqrt{2 \pi} e^{2 z^2} {\rm erf}(\sqrt{2} z) - 4 z \right) } \times \nonumber \\
		&  \left( \sqrt{\pi} y e^{-(y-4 i z)^2/8} \left[ {\rm erf}\left( \frac{4 z - i y}{2\sqrt{2}}\right)  + {\rm erf}\left( \frac{4 z + i y}{2\sqrt{2}} \right) \right]   \right. \nonumber \\
		& \left. + 2 \sqrt{2} i \left(-1 + e^{2 i y z}\right) \right) ,
	\end{align}
	where $y \equiv \Delta x k_0 = \Delta x m_{A'} v_0$. The full result can be simplified using Taylor expansion over the large $v_{\rm esc}$ or $z$,  which leads to the leading term and the next-to-leading term as
	\begin{align}
		&	\mathcal{S}(\Delta \vec{x})  \approx e^{- \frac{1}{8}y^2} - \frac{2\sqrt{2}}{\sqrt{\pi} y} e^{-2 z^2} \sin(y z).
	\end{align}
	It is clear that an exponential factor further suppresses the next-to-leading term compared with the leading term. Therefore, the leading term is already a good approximation in the numeric calculation. This result agrees with the equal-time two-point correlation function in Ref.~\cite{Derevianko:2016vpm}. Our method utilizes the random phase of each $ k$ mode, which is equivalent to the specific Fock state in Ref.~\cite{Derevianko:2016vpm}.

	The above results are derived based on the Standard Halo Model. However, there are possibilities for a non-standard halo model and non-trivial substructure like the streams S1/S2, which can significantly impact the signal~\cite{Foster:2017hbq, OHare:2018trr, Evans:2018bqy, OHare:2019qxc}. One aspect of the impact is the modification of the shape of the signal power spectral density, which can be noticed with high-frequency resolution measurements such as the axion resonant cavity searches. For the dark photon signal with radio telescope searches, the frequency resolutions for existing telescopes are $7.63$ kHz for FAST, $700$ Hz for LOFAR, while for future telescopes, are $1$ kHz for SKA1-Low and $200$ Hz for SKA1-Mid. Compared with their operating frequency ranges, only SKA1-Mid can have a chance to resolve the structure of signal power spectral density. Therefore, the current radio telescopes are challenging to discriminate different halo models. 
	
	Another aspect of the impact is the change in correlation patterns. Eq.~(\ref{eq:EfieldinSHM}) should be implemented with the new velocity distribution function, thus modifying the form of the correlation $\mathcal{S}(\Delta \vec{x})$. The network of detectors can reveal the correlation pattern and the directional information of the DM phase space distribution~\cite{Foster:2020fln}. 
	While constructing additional detectors for axion experiments to form a network can be challenging, it is comparatively simple for antenna arrays in dark photon detection since each antenna has a straightforward structure that is easy to build. For example, LOFAR, which consists of 40 stations with multiple antennas inside each station, has already been built. Although increasing data storage and modifying data processing for local detection via radio telescopes present additional challenges, today's technology provides the means to overcome these challenges. 
	
	\section{Analysis of the FAST data} 
	\label{sec:FAST-data}
	
	In this part, we present the detailed analysis of the FAST data to calculate the upper limit on the mixing parameter $\epsilon$ in the DPDM model. 
	
	The FAST observation was conducted between 2020-12-14 at 07:00:00 and 2020-12-14 at 08:50:00.
	The 19-beam receiver equipped on the FAST~\cite{Jiang:2019rnj} recorded data of two polarisations through 65536 spectral channels, covering the frequency range of $1-1.5$ GHz. The SPEC(W+N) backend was employed in the observation, with 1 second sampling time. The  ``ON/OFF'' observation mode of the FAST was used for the original motivation, constraining the WIMP property by searching for synchrotron emission in the dwarf spheroidal galaxy Coma Berenices. In each round of observation, the central beam of the FAST was pointed to the Coma Berenices for 350 seconds, and the central beam of the FAST was moved to the ``OFF source" position, which was a half degree away from the ``ON source" without known radio sources, for another 350 seconds. There were nine rounds of ON/OFF observation. The low noise injection mode was used for signal calibration, with the characteristic noise temperature of about 1.1 K~\cite{2020RAA....20...64J}. In the whole observation, the noise diode was continuously
	switched on and off for a period of 1 second.
	
	Let $P^{\rm cal_{\rm off}}$ be the original instrument reading without noise injection, $P^{\rm cal_{\rm on}}$ be the reading with noise injected. Then the system temperature $T_{\rm sys}$ can be calibrated as~\cite{2020RAA....20...64J}
	\begin{equation}\label{Temperature calibration}
		T_{\rm sys}=\frac{P^{\rm cal_{\rm off}}}{P^{\rm cal_{\rm on}}-P^{\rm cal_{\rm off}}}T_{\rm noise},
	\end{equation}
	where $T_{\rm noise}$ is the pre-determined noise temperature measured with hot loads.
	$T$ from both polarizations were checked without large variation, and each pair of polarization temperatures was added to
	get the average. Then, the polarization-averaged $T_{\rm sys}$ was converted to flux density with pre-measured antenna gain, which is beam dependent~\cite{2020RAA....20...64J}. With these procedures, the flux density for each beam, each time bin, and each frequency bin could be obtained. Finally, these flux densities in the time domain are checked, and time bins, when the telescope is not stable due to switching between ON and OFF, were masked, leaving 2232 time bins (1116 time bins for the ON observations and 1116 time bins for the OFF observations) for each frequency.

	\subsection{Data processing}
	We are going to utilize the ON data. As mentioned above, each of the 19 beams covers the radio frequency range 1$-$1.5~GHz divided evenly into 65536 frequency bins (the bandwidth is $\approx 7.63~\kHz$), and each frequency bin has 1116 time bins. The data suffer from large noise fluctuation [\eg radio frequency interference (RFI)]. To reduce the effect of the noise fluctuation, we adopt the following method to clean the data (see \eg Ref.~\cite{Foster:2022fxn} for a similar method but more complicated in their case). 
	
	The cleaning process is applied to the time series at each frequency bin. For each frequency bin $i$, we divide the 1116 time bins consecutively into 33 groups, and thus each group contains 33 time bins (the last group contains 27 time bins). We identify the group with the smallest variance, $\sigma_{\rm ref}^2$, as the reference group. The mean value of the reference group is labeled as $\mu_{\rm ref}$.
	We then regroup the time series by keeping only the bins with deviation from the reference group smaller than $3\sigma_{\rm ref}$ (\ie $|O - \mu_{\rm ref}| < 3 \sigma_{\rm ref}$). 
	
\begin{figure}
		\centering
		\includegraphics[width=1\linewidth]{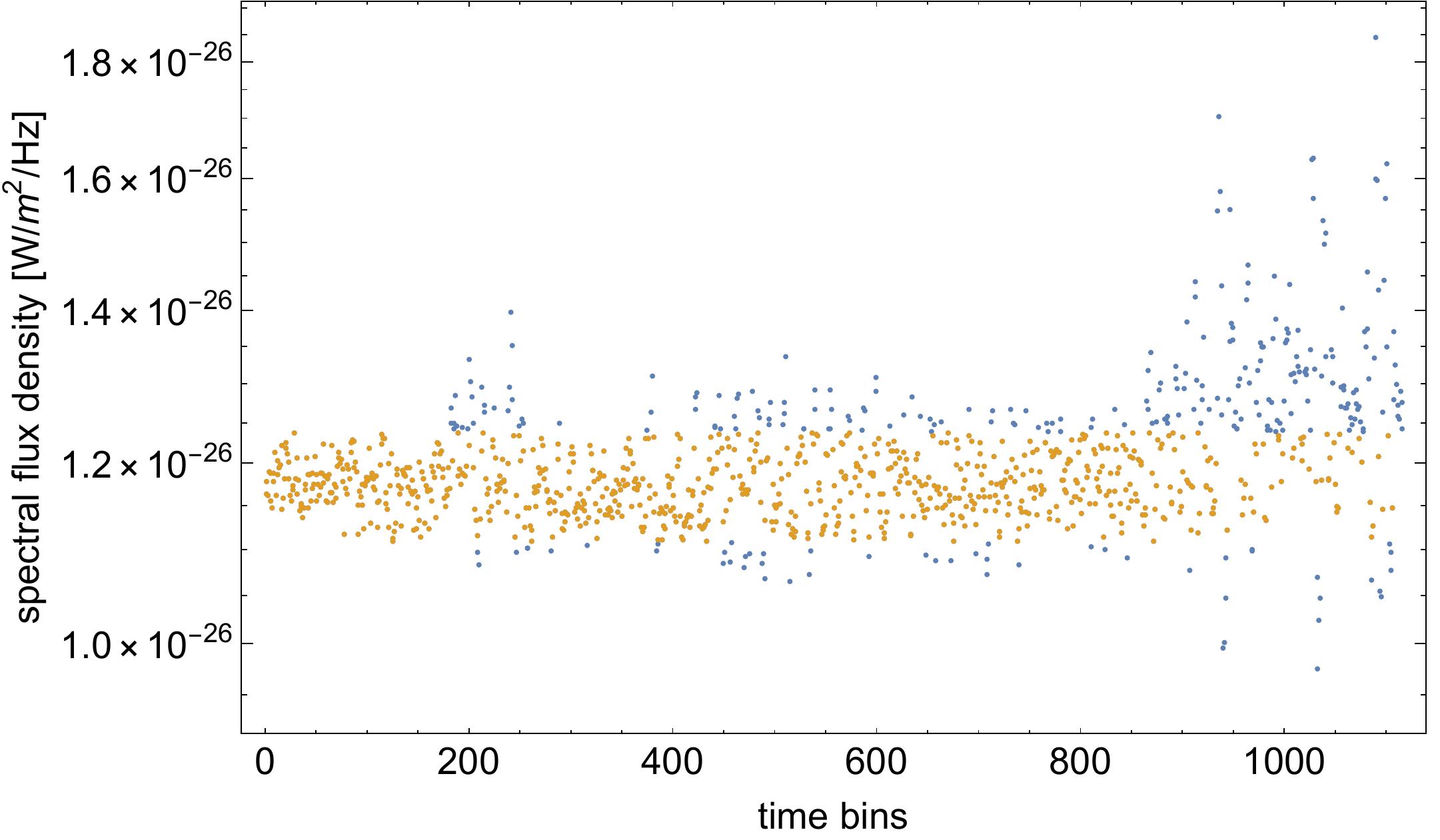}
		\caption{Process of data cleaning. This is for frequency bin $i=20000$ ($f\approx 1.15258~\GHz$) of beam 1 as an example. The blue points are the excluded data, while the yellow points are the remaining data after the cleaning process.}
		\label{fig:data_filtering}
	\end{figure}

\begin{figure}
		\centering
		\includegraphics[width=1\linewidth]{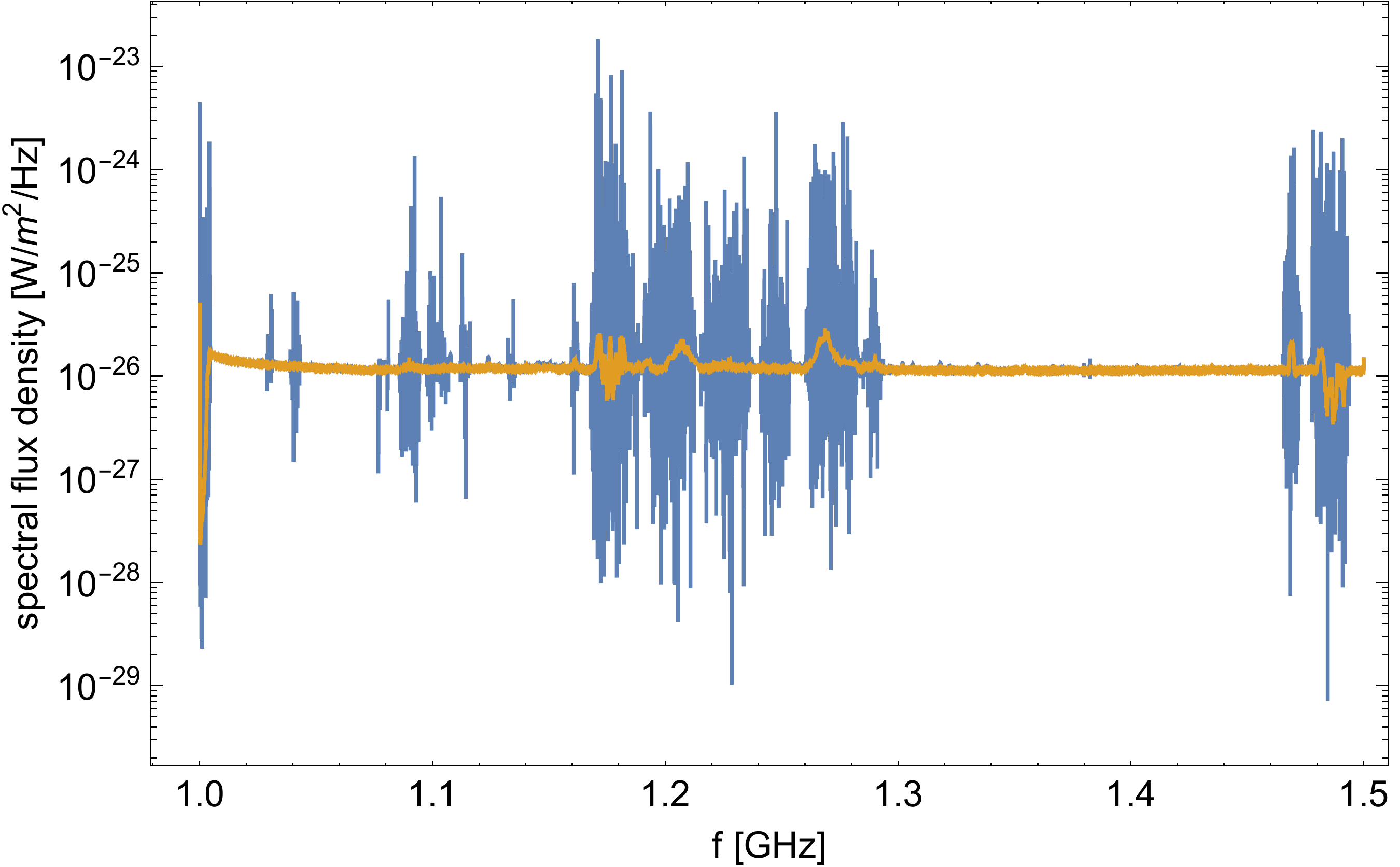}
		\caption{For all frequency bins, we have applied the data cleaning process. This plot summarizes the data (mean value) at all frequency bins of beam 1 as an example. The blue curve is the original raw data, while the yellow curve is the result ($\bar{O}_i$) after the cleaning process.}
		\label{fig:data_filtering_all}
	\end{figure}

	After the above cleaning process, the number of the remaining time bins in the time series reduces to $N_i$ with the mean $\bar{O}_i$,
	\bea\label{eq:sde_mean}
	\bar{O}_i = \frac{1}{N_{i}}\sum_{j=1}^{j=N_i}O_{i,j} \ ,
	\eea
	where $i$ and $j$ label the frequency and time, respectively. 
	The standard deviation of the mean (called the standard error, $\sigma_{\bar{O}_i}$) is thus
	\beq\label{eq:sderror_mean}
	\sigma_{\bar{O}_i}^2 =
	{\rm Var}(\bar{O}_i)
	=\frac{1}{N_{i}(N_{i}-1)}\sum_{j=1}^{j=N_i}(O_{i,j}-\bar{O}_i)^2 \ ,
	\eeq
	which can be used as statistical uncertainty. As an example, in Fig.~\ref{fig:data_filtering}, we present the time series before (yellow$+$blue) and after (yellow) cleaning for the frequency bin $i=20000$ ($f\approx 1.15258~\GHz$) of beam 1. For each beam, we apply the cleaning process and calculate $\bar{O}_i$ and $\sigma_{\bar{O}_i}$ for all the 65536 frequency bins. The filtered time-average brightness distribution as a function of frequency is shown by the yellow curve in Fig.~\ref{fig:data_filtering_all}, compared with the time-average brightness without data cleaning (blue curve). 
	
	\subsection{Background fit}
	We fit the background around the frequency bin $i_0$ with a polynomial function
	\beq\label{eq:bg_fit}
	B(a, f) = a_n f^n + a_{n-1} f^{n-1} +...+ a_1 f + a_0. 
	\eeq
	It describes the observed data $\bar{O}_i$ locally from the frequency bin $i=i_0-k$ to the bin $i=i_0+k$. In practice, we choose $k=5$ and $n=3$. 
	
	The systematic uncertainty due to the background fit can be estimated as follows. We first use the method of weighted least squares to fit the background by minimizing the function 
	\beq\label{eq:w_least_squares}
	\sum^{i_0+k}_{i=i_0 - k} \left(\frac{B(a, f_i)-\bar{O}_i}{\sigma_{\bar{O}_i}}\right)^2
	,~~
	\text{(bin $i_0$ excluded).}
	\eeq
	The result is labeled as $\tilde{a}$ so that the background $B(\tilde{a}, f)$ minimizes \eqref{eq:w_least_squares}. Then the systematic uncertainty on bin $i_0$ can be estimated by the deviation of the data to the fitted background curve defined as
	\beq
	(\sigma_{i_0}^{\rm sys})^2 & = {\rm Var}(\delta)= \frac{1}{2k-1}\sum_{i=i_0-k}^{{i_0+k}}(\delta_{i}-\bar{\delta})^2,\\
	&
	\text{(bin $i_0$ excluded).}
	\eeq
	where 
	\beq
	\delta_i \equiv B(\tilde{a},f_i)-\bar{O}_i.
	\eeq
	We then add the statistical uncertainty and systematic uncertainty in quadrature form as the total uncertainty of the frequency bin $i_0$,
	\beq\label{eq:sigma_tot}
	(\sigma_{i_0}^{\rm tot})^2 = \sigma_{\bar{O}_{i0}}^2 + (\sigma_{i_0}^{\rm sys})^2.
	\eeq
	Repeating this process, we can get the total uncertainties for all frequency bins for all beams.
	
	\begin{figure}
		\centering
		\includegraphics[width=1\linewidth]{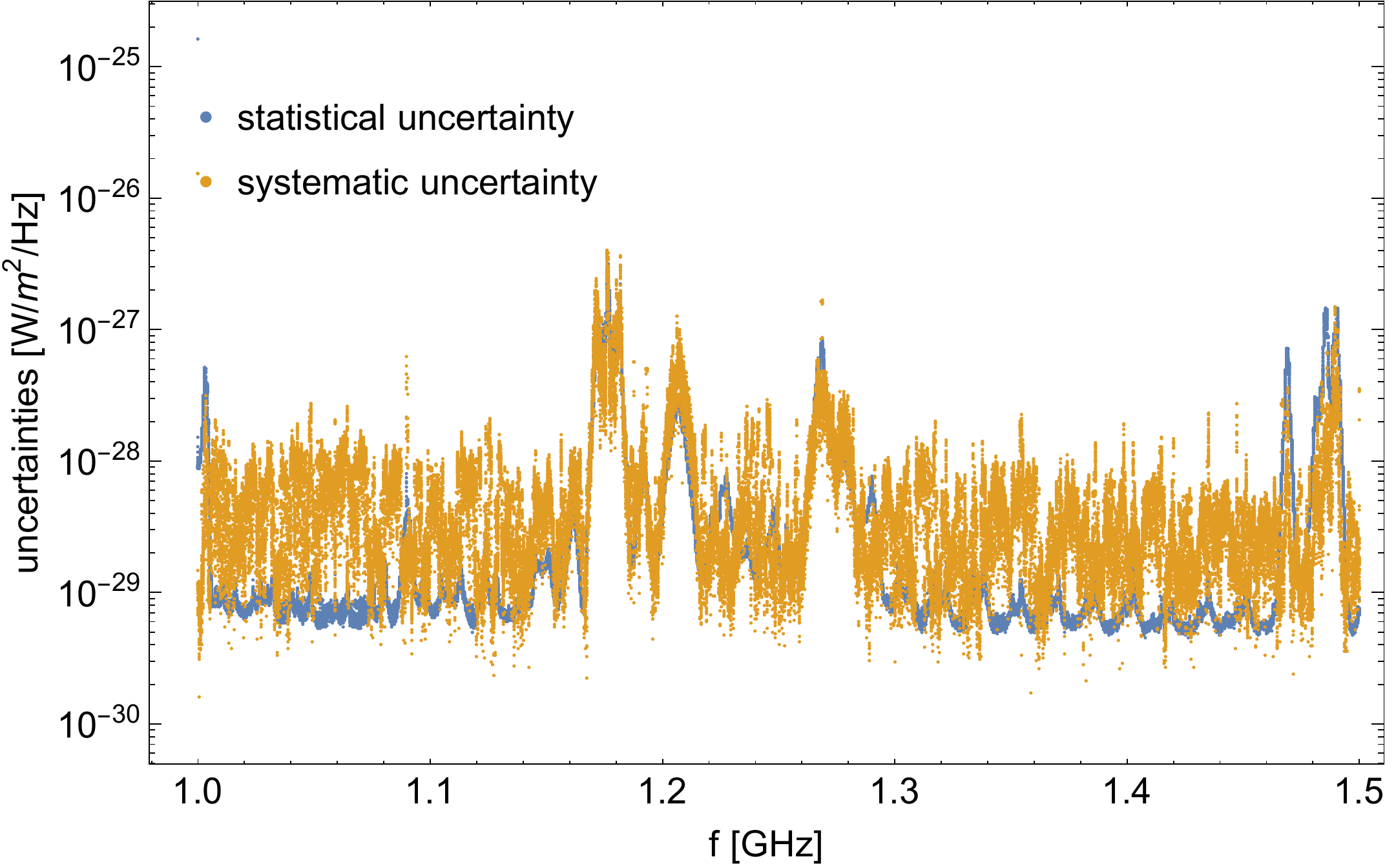}
		\caption{Statistical uncertainties $\sigma_{\bar{O}_{i0}}$ and systematic uncertainties $\sigma_{i_0}^{\rm sys}$ for all frequency bins in the range $1-1.5$~GHz of beam 1 as an example.}
		\label{fig:uncertainties}
	\end{figure}
	
	In Fig.~\ref{fig:uncertainties}, we show the statistical and systematic uncertainties for all frequency bins in the range $1-1.5$~GHz. This is plotted using the data of beam 1 as an example.

	\subsection{Likelihood-based statistical test}
	In this part, we adopt the likelihood-based statistical method~\cite{Cowan:2010js} to set upper limits on the mixing parameter $\epsilon$ in the DPDM model. To calculate the likelihood of the DPDM-induced EM equivalent spectral flux density $S$ at the frequency bin $i_0$, we construct the likelihood function using $k$ neighboring bins on both sides of $i_0$, 
	\beq\label{eq:likelihood}
	& L(S, a) = 
	\\ 
	&\prod_{i=i_0 - k}^{i_0+k} \frac{1}{\sigma_{i}^{\rm tot} \sqrt{2\pi}}\exp{\left[-\frac{1}{2}\left( 
		\frac{B(a, f_i) + S \delta_{ii_0} - \bar{O}_i}{\sigma_{i}^{\rm tot}}
		\right)^2\right]} \ .
	\eeq
	Here $\bar{O}_i$ and $\sigma_{i}^{\rm tot}$ are calculated from Eqs.~\eqref{eq:sde_mean} and \eqref{eq:sigma_tot}, respectively. $B(a, f_i)$ is the background function defined in \eqref{eq:bg_fit}, where the coefficients $a=(a_n, a_{n-1},...,a_0)$ are treated as nuisance parameters. 
	
	To set the \textit{one-sided} upper limit on the dark photon parameter $\epsilon$, we use the following test statistic~\cite{Cowan:2010js}
	\beq\label{eq:test-statistic-q}
	q_S = 
	\begin{cases}
		-2\ln\left[\frac{L(S, \hat{\hat{a}})}{L(\hat{S}, \hat{a})}\right],~~~ \hat{S}\leq S \\
		0,~~~~~~~~~~~~~~~~~~~~ \hat{S}>S
	\end{cases}.
	\eeq
	In the denominator, $\hat{S}$ and $\hat{a}$ denote the values of the signal and nuisance parameters at which the likelihood gets maximized. In the numerator, $\hat{\hat{a}}$ denotes the values of the nuisance parameters at which the likelihood gets maximized for a specified value of $S$. We see that the test statistic $q_S$ is a function of $S$ and $q_S\geq 0$.  
	
	Applying the Wald approximation here, we have (see \eg Ref.~\cite{Cowan:2010js}) 
	\beq\label{eq:likelihood-ratio-Wald}
	-2\ln\left[\frac{L(S, \hat{\hat{a}})}{L(\hat{S}, \hat{a})}\right] \simeq \frac{(S-\hat{S})^2}{\sigma^2}
	\eeq
	It can be proved that in our case where the nuisance parameters and signal are defined as in \eqref{eq:bg_fit} and \eqref{eq:likelihood}, the relation \eqref{eq:likelihood-ratio-Wald} is exact. Furthermore, $\sigma \sim \sigma_{i}^{\rm tot}$ is a constant which does not change with $S$. Here $\hat{S}$ follows the normal distribution with the mean $S'$ (the assumed true signal strength) and the standard derivation $\sigma$~\cite{Cowan:2010js}. Thus, \eqref{eq:likelihood-ratio-Wald}  follows the $\chi^2$ distribution for $1$ degree of freedom if we choose $S'=S$. Then, it can be demonstrated that our test statistic \eqref{eq:test-statistic-q} follows the distribution:
	\beq
	f(q_S|S)= \frac{1}{2}\delta(q_S) + \frac{1}{2}\frac{1}{\sqrt{2\pi}}\frac{1}{\sqrt{q_S}}{\rm e}^{-q_S/2}
	\eeq
	which is named as the half-$\chi^2$ distribution in Ref.~\cite{Cowan:2010js}. The cumulative distribution is $F(q_S|S) = \Phi(\sqrt{q_S})$ where $\Phi(x)$ is the cumulative distribution function of the standard normal distribution with mean=0 and variance=1. We define the $p$-value as a measurement of how far the assumed signal $S$ is from the null $S=0$,
	\beq\label{eq:p_mu_ShatP}
	p_S = [1-\Phi(\sqrt{q_S})]/[1-\Phi(\sqrt{q_0})].
	\eeq
	In practice, one usually sets $p_S=0.05$. The corresponding value of $S$ is denoted as $S_{\rm lim}$. Any $S>S_{\rm lim}$ is excluded at $95\%$ confidence level (C.L.). $S_{\rm lim}$ is thus called the one-sided upper limit. 
	
	In the previous section, we have numerically simulated the equivalent EM flux density induced by DMDP on FAST at different frequencies, $I_{\rm FAST}^{\rm eqv}$; see \eqref{eq:I_FAST_eqv} and Fig.~\ref{fig:CFAST}. Divided by the bandwidth of FAST data, we can further get the spectral flux density $S^{\rm eqv}_{\rm FAST}\equiv I^{\rm eqv}_{\rm FAST}/\mathcal{B}$. Then the upper limit on $\epsilon$ can be obtained using the relation $S_{\rm lim} = S^{\rm eqv}_{\rm FAST}$. Finally, we apply the process of finding $S_{\rm lim}$ to all frequency bins of one of the total 19 beams. Then we can get the constraints on $\epsilon$ in the frequency range 1$-$1.5~GHz from that beam. 
	
	Due to velocity dispersion, the bandwidth of DPDM is $\sim m_{A'}v_0^2/2\pi \sim 0.15{\rm~kHz}\times (m_{A'}/\mu{\rm eV})$, which is smaller than the bandwidth ($\approx 7.63$ kHz) of the FAST data in the range 1$-$1.5~GHz. For DPDM signal falls into one frequency bin, the limits are the same for all frequencies in this bin. However, there is a special case that the DPDM signal sits right at the edge of bins and contributes to the two adjacent bins due to the broadening. Therefore, one expects the limits to become weaker because the backgrounds from both bins contribute simultaneously. Usually, this special case can be avoided if one can re-bin the data. However, the bins from FAST data are fixed already. Therefore, in this case, we merge the two bins and repeat the above procedure to find $S_{\rm lim}$. In general, we expect the constraints on $\epsilon$ would be weakened by a factor of about $\sim\sqrt{2}$ but is subject to changes in statistical and systematic uncertainties. 
	
\begin{figure}[htp]
		\centering
		\includegraphics[width=1\linewidth]{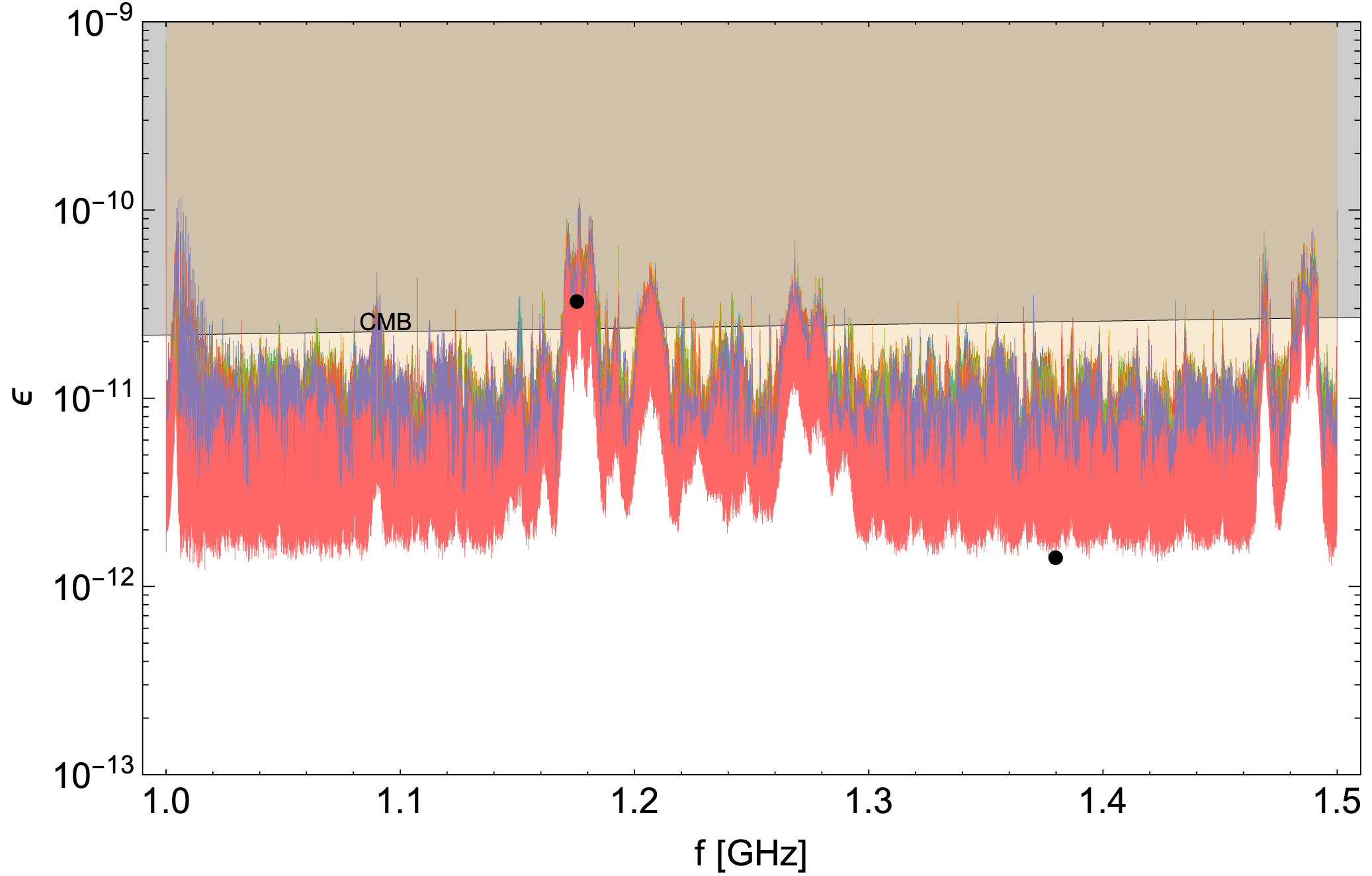}
		\caption{The summary of the constraints on $\epsilon$ from the total 19 beams of FAST. Different colors represent constraints from different beams. The red curve is the final result which is obtained by selecting the strongest one of all 19 constraints at each frequency bin. In comparison, we also show the existing constraint from CMB~\cite{Arias:2012az}. }
		\label{fig:epsilon_all_beams}
	\end{figure}

	We repeat the above steps for all beams and get 19 sets of constraints, which are similar to each other since the beams recorded similar data. Now for each frequency bin, there are 19 similar limits, and we choose the strongest one as the final limit at that frequency bin. In addition, we show the 19 sets of constraints in Fig.~\ref{fig:epsilon_all_beams},  and the red curve is the final result. We emphasize that every single frequency in the 1$-$1.5 GHz is constrained by the actual FAST data without any extrapolation because the signal can be contained in one bandwidth. In Fig.~\ref{fig:epsilon_all_beams}, we have plotted the full 65536 frequency bins, which is already quite busy to show. Therefore, to improve the readability, we took the average of every four bins in the final plot in the main text to better illustrate the envelope of the constraints.
	
	\begin{figure}[htb]
		\centering
		\includegraphics[width=1\linewidth]{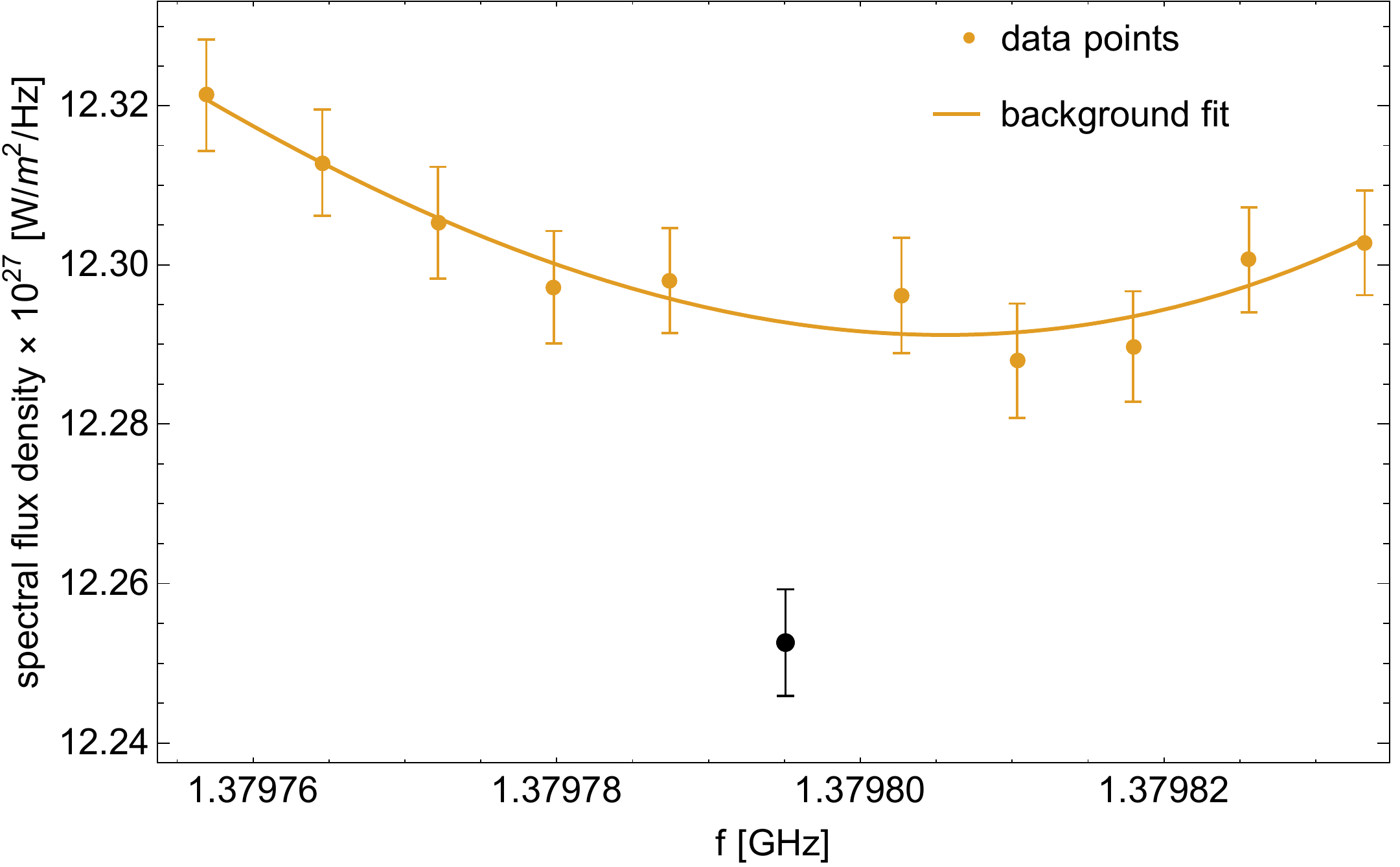}
		\caption{The black dot is the data $\bar{O}_i$ of the frequency bin that we choose from Fig.~\ref{fig:epsilon_all_beams} as an example to show how we get a relatively good constraint.
			It is bin 49781 ($f=1.379795$~GHz) of beam 14. The solid curve is the background fit to the neighboring $2k=10$ data points; see \eqref{eq:w_least_squares}. The error bar is the value of statistical uncertainty associated with each data point.}
		\label{fig:point_good}
	\end{figure}
	
	\begin{figure}[htb]
		\centering
		\includegraphics[width=1\linewidth]{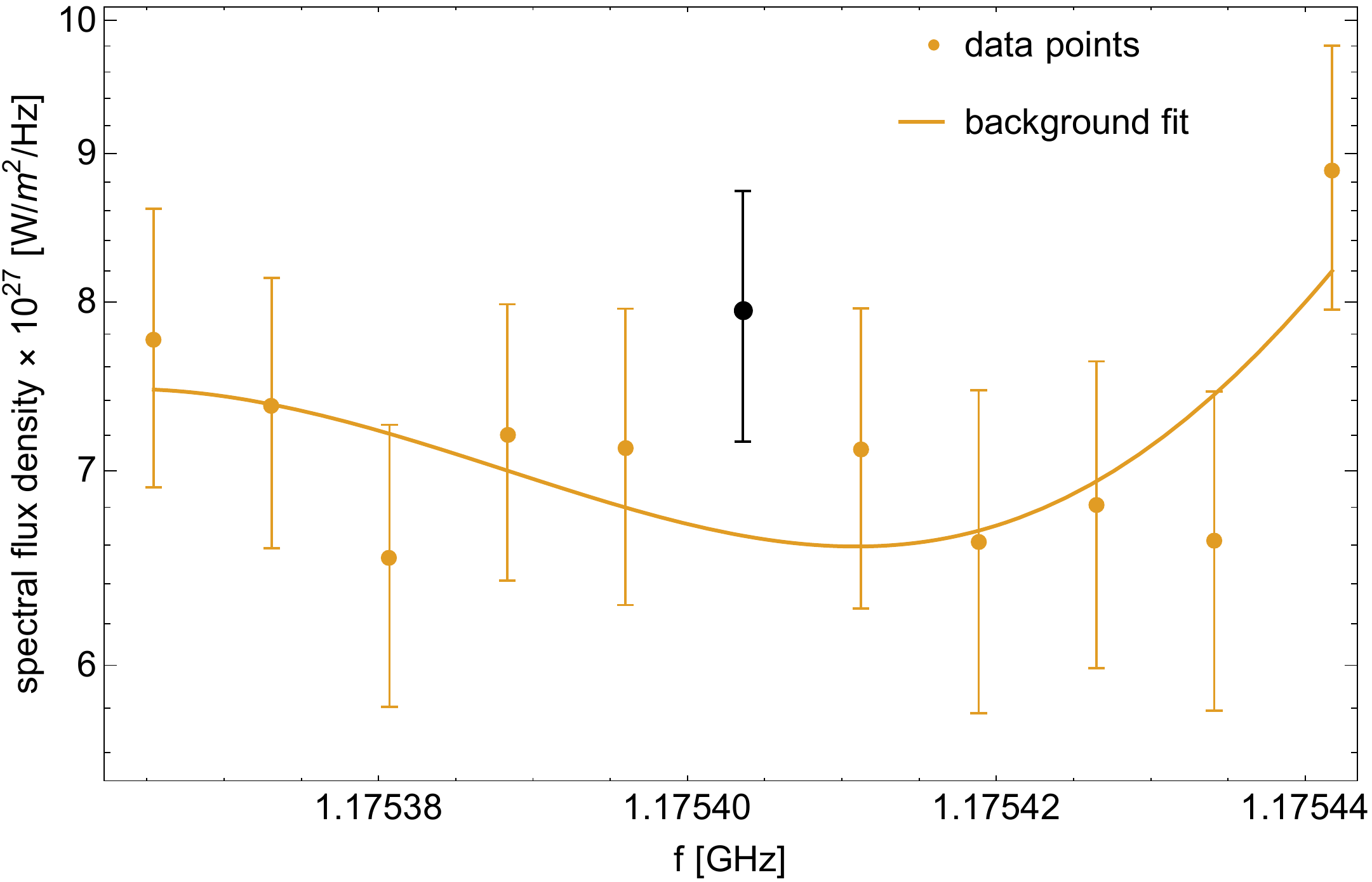}
		\caption{Same as Fig.~\ref{fig:point_good} but to show how we get a relatively bad constraint. The black dot corresponds to the bin 22991 ($f=1.175404$~GHz) of beam 1.}
		\label{fig:point_bad}
	\end{figure}
	
	From Fig.~\ref{fig:epsilon_all_beams}, we see that for some frequency bins, the constraints can reach $\epsilon\sim 10^{-12}$, while for some other frequencies, the constraints are only $\epsilon\sim 10^{-11}$. The difference between them is due to the quality of observed data. We are going to demonstrate this more clearly in the following text. We choose one frequency bin from Fig.~\ref{fig:epsilon_all_beams} where the constraint is ``good'' [bin 49781 ($f=1.379795$~GHz) of beam 14 with
	$\epsilon \approx 1.4\times 10^{-12}$] and another frequency bin where the constraint is ``bad'' [bin 22991 ($f=1.175404$~GHz) of beam 1 with
	$\epsilon \approx 3.3\times 10^{-11}$] as examples to show how the data quality affects the limits. The two examples are denoted as black dots in Fig.~\ref{fig:epsilon_all_beams}. In Fig.~\ref{fig:point_good} and Fig.~\ref{fig:point_bad}, we respectively plot the $\bar{O}_i$ value of the two examples (denoted as black dots) with the neighboring $2k=10$ data points. The error bar is the value of statistical uncertainty associated with each data point. The solid curve is the background fit from \eqref{eq:w_least_squares}. Fig.~\ref{fig:point_good} shows a relatively small fluctuation and the example good point has the uncertainties $\sigma_{\bar{O}_{i0}}\approx 6.7\times 10^{-30} ~{\rm W/m^2/Hz}$, $\sigma_{i_0}^{\rm sys}\approx 2.9\times 10^{-30}~{\rm W/m^2/Hz}$ and $\sigma_{i_0}^{\rm tot}\approx 7.3\times 10^{-30}{\rm W/m^2/Hz}$; while Fig.~\ref{fig:point_bad} shows a relatively large fluctuation and the example bad point has the uncertainties $\sigma_{\bar{O}_{i0}}\approx 7.9\times 10^{-28} ~{\rm W/m^2/Hz}$, $\sigma_{i_0}^{\rm sys}\approx 4.8\times 10^{-28}~{\rm W/m^2/Hz}$ and $\sigma_{i_0}^{\rm tot}\approx 9.2\times 10^{-28}{\rm W/m^2/Hz}$. We see that the former case has a relatively small total uncertainty, which is why it has a better constraint. In addition, maximizing the likelihood~\eqref{eq:likelihood} gives a negative best-fit $\hat{S}$ for the former case, which also contributes to finally obtaining a better constraint.

	\bibliographystyle{utphys}
	\bibliography{ref}

\begin{thebibliography}{100}
\bibitem{footnote}
The SKA telescope will be constructed in two phases: SKA1 is being designed
  now; SKA2 is planned to follow, but not yet fully defined. Therefore, in this
  work we only calculate the SKA1 sensitivity for DPDM.
\end{thebibliography}

\end{document}